\documentclass[floats,floatfix,showpacs,amssymb,prd,twocolumn,superscriptaddress,nofootinbib,nolongbibliography,reprint,preprintnumbers]{revtex4-1}

\usepackage{amssymb,amsmath,verbatim,mathtools,needspace,enumitem,etoolbox,graphicx,physics,microtype,afterpage,xspace,tabularx,lmodern,multirow,bm}
\usepackage{textcomp, gensymb}

\usepackage{braket}

\usepackage{relsize}

\usepackage{dcolumn}

\usepackage[normalem]{ulem}
\usepackage{placeins}
\usepackage{comment}
\usepackage[dvipsnames, usenames]{xcolor}
\usepackage{colortbl}
\definecolor{linkcolor}{rgb}{0.0,0.3,0.5}
\usepackage[unicode, colorlinks=true, linkcolor=linkcolor, citecolor=linkcolor, filecolor=linkcolor, urlcolor=linkcolor, linktocpage, breaklinks]{hyperref}
\usepackage[all]{hypcap}
\usepackage[T1]{fontenc}
\usepackage[utf8]{inputenc}
\usepackage{mathrsfs}
\usepackage[usenames,dvipsnames]{xcolor}
\hypersetup{colorlinks=true,citecolor=romared,linkcolor=romared,urlcolor=romared}

\definecolor{romared}{RGB}{142,0,28}

\newcommand{\be}{\begin{equation}}
\newcommand{\ee}{\end{equation}}

\def\be{\begin{equation}}
\def\ee{\end{equation}}
\newcommand{\beq}{\begin{eqnarray}}
\newcommand{\eeq}{\end{eqnarray}}

\usepackage{aas_macros}
\usepackage{makecell}
\usepackage{soul}
\usepackage[nolist,nohyperlinks]{acronym}
\usepackage{orcidlink}
\usepackage{lipsum}
\usepackage{booktabs}

\acrodef{LSC}[LSC]{LIGO Scientific Collaboration}
\acrodef{BH}{black hole}
\acrodef{NS}{neutron star}
\acrodef{PN}{Post-Newtonian}
\acrodef{BBH}{binary black-hole}
\acrodef{BNS}{binary neutron-star}
\acrodef{NSBH}{neutron-star black-hole}
\acrodef{NR}{numerical relativity}
\acrodef{GW}{gravitational wave}
\acrodef{PSD}{power spectral density}
\acrodef{aLIGO}{Advanced Laser interferometer Gravitational-Wave Observatory}
\acrodef{AZDHP}{aLIGO zero detuned high power density}
\acrodef{GR}{general relativity}
\acrodef{PE}{parameter estimation}
\acrodef{LAL}{LIGO algorithm library}
\acrodef{TPI}{tensor-product interpolant}
\acrodef{SVD}{singular value decomposition}
\acrodef{SNR}{signal-to-noise ratio}
\acrodef{ODE}{ordinary differential equation}
\acrodef{PDE}{partial differential equation}
\acrodef{ROM}{reduced order model}
\acrodef{QNM}{quasi-normal mode}
\acrodef{IMR}{inspiral-merger-ringdown}
\acrodef{LVK}{LIGO-Virgo-KAGRA}
\acrodef{SXS}{Simulating eXtreme Spacetimes}

\newcommand{\ias}{\affiliation{School of Natural Sciences, Institute for Advanced Study, 1 Einstein Drive, Princeton, NJ 08540, USA}}

\newcommand{\utexas}{\affiliation{Center for Gravitational Physics, University of Texas at Austin, Austin, TX 78712, USA}}

\newcommand{\ucsb}{\affiliation{Department of Physics, University of California, Santa Barbara, CA 93106, USA}}
\newcommand{\icts}{\affiliation{International Centre for Theoretical Sciences, Tata Institute of Fundamental Research, Bangalore 560089, India}}
\newcommand{\kicp}{\affiliation{Kavli Institute for Cosmological Physics, The University of Chicago, 5640 South Ellis Avenue, Chicago, IL 60637, USA}}
\newcommand{\cmi}{\affiliation{Chennai Mathematical Institute, Siruseri, Tamil Nadu 603103, India}}

\newcommand{\orcid}[1]{\href{https://orcid.org/#1}{\includegraphics[width=10pt]{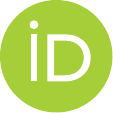}}}
\newcommand{\ben}{\begin{enumerate}}
\newcommand{\een}{\end{enumerate}}

\def\be{\begin{equation}}
\def\ee{\end{equation}}
\def\beq{\begin{eqnarray}}
\def\eeq{\end{eqnarray}}

\def\apjl{\rm{ApJ}}
\def\apjs{\rm{ApJS}}
\def\aap{\rm{A\&A}}

\graphicspath{{./Plots/}}

\allowdisplaybreaks

\newcommand{\PMuninfodds}{3.17^{+0.14}_{-0.09}}
\newcommand{\PMgwtcodds}{5.41^{+0.34}_{-0.47}}
\newcommand{\PMfPBHopt}{-1.55^{+0.28}_{-0.45}}
\newcommand{\PMfBHopt}{2.02^{+0.34}_{-0.44}}
\newcommand{\PMIMBHcluster}{-5.12^{+0.37}_{-0.45}}
\newcommand{\gSISuninfodds}{2.98^{+0.10}_{-0.13}}
\newcommand{\gSISgwtcodds}{5.84^{+0.63}_{-0.41}}
\newcommand{\gSISfone}{1.24^{+0.59}_{-0.39}}
\newcommand{\gSISfm}{0.24^{+0.54}_{-0.37}}
\newcommand{\CISuninfodds}{2.04^{+0.12}_{-0.11}}
\newcommand{\CISgwtcodds}{5.38^{+0.59}_{-0.50}}
\newcommand{\CISfone}{0.72^{+0.56}_{-0.44}}
\newcommand{\CISfm}{-0.29^{+0.49}_{-0.41}}
\newcommand{\MoptfPBH}{627}
\newcommand{\MoptfBH}{938}

\newcommand{\Punlensed}{0.13^{+0.44}_{-0.07}}

\newcommand{\PClogfPBHopt}{-6.26^{+0.51}_{-1.35}}
\newcommand{\PClogfBHopt}{-3.10^{+0.62}_{-1.11}}
\newcommand{\PCloggSISfone}{-3.20^{+0.66}_{-1.09}}
\newcommand{\PCloggSISfm}{-4.05^{+0.67}_{-1.15}}
\newcommand{\PClogCISfone}{-3.04^{+0.63}_{-0.93}}
\newcommand{\PClogCISfm}{-4.07^{+0.45}_{-0.89}}
\newcommand{\PClogIMBHcluster}{-10.08^{+1.03}_{-1.14}}

\begin{document}

\pagenumbering{arabic}

\title{The diffraction-lensing interpretation of GW231123 with astrophysical priors}

\author{Mark Ho-Yeuk Cheung \orcid{0000-0002-7767-3428}}
\email{mcheung@ias.edu}
\ias

\author{Digvijay Wadekar
\orcid{0000-0002-2544-7533}}
\utexas

\author{Matias~Zaldarriaga~\orcid{0009-0007-8315-6703}}
\ias

\author{Tejaswi Venumadhav
\orcid{0000-0002-1661-2138}}
\ucsb
\icts

\author{Javier Roulet
\orcid{0000-0003-3268-4796}}
\ias
\kicp

\author{Ajit Kumar Mehta
\orcid{0000-0002-7351-6724}}
\cmi
\icts

\date{\today}

\begin{abstract}
GW231123, if unlensed, is a rare binary black hole merger with high masses and high spins for both progenitors.
We show that the signal is better fitted by a lower-mass, lower-spin merger that is diffraction-lensed by an isolated object of redshifted mass $\sim 1000\,\rm M_\odot$, modeled either as a point mass or as a spherically symmetric compact halo.
Because diffraction-lensed events are also rare, hypothesis testing should quote the posterior odds ratio rather than the Bayes factor, which requires quantifying our prior belief in the two hypotheses.
We adopt the GWTC-5 population distribution as our source parameter prior, and quantify the prior on the lens hypothesis through the lensing optical depth.
For a point-mass lens, observational constraints on the abundance of black holes in the Universe yield an upper bound on the optical depth, and hence on the posterior odds, which do not rule out lensing.
However, using a predicted mass function of intermediate-mass black holes formed in star clusters gives a low optical depth that strongly disfavors lensing.
For a dark matter halo lens, standard collisionless cold dark matter does not form halos compact enough to give the required optical depth, but self-interacting dark matter with a large cross section at low velocities can trigger gravothermal collapse and form them, in which case the posterior odds are inconclusive.
In either case, from a frequentist perspective, we show that detecting a lensed event with properties like GW231123 is unlikely.
These conclusions apply to isolated lenses, while a lens embedded in an external gravitational potential could change the picture.
\end{abstract}

\preprint{000000}

\maketitle

\section{\label{sec:intro}Introduction}

\begin{table*}
\centering
\renewcommand{\arraystretch}{1.5}
\begin{tabular}{lllll}
\toprule
Lens model & Source prior & Lens prior & $\log_{10}\mathcal{O}^{\rm L}_{\rm unL}$ & $\log_{10} p^{\rm L}_{{\rm obs}\geq1}$ \\
\midrule
\addlinespace[2pt]
\multirow{5}{*}{PM} & Uninformative & Uninformative & $\PMuninfodds$ & --- \\
 & GWTC-5 & Uninformative & $\PMgwtcodds$ & --- \\
 & GWTC-5 & $f_{\rm PBH}$ constraint, optimal mass& $\PMfPBHopt$ & $\PClogfPBHopt$ \\
 & GWTC-5 & $f_{\rm BH}$ constraint, optimal mass& $\PMfBHopt$ & $\PClogfBHopt$ \\
\rowcolor{gray!15}[\tabcolsep][\tabcolsep] \cellcolor{white} & GWTC-5 & IMBHs formed in star clusters & $\PMIMBHcluster$ & $\PClogIMBHcluster$ \\
\midrule
\addlinespace[2pt]
\multirow{4}{*}{gSIS} & Uninformative & Uninformative & $\gSISuninfodds$ & --- \\
 & GWTC-5 & Uninformative & $\gSISgwtcodds$ & --- \\
 & GWTC-5 & SIDM collapsed ($f_{\rm coll}=1$) & $\gSISfone$ & $\PCloggSISfone$ \\
\rowcolor{gray!15}[\tabcolsep][\tabcolsep] \cellcolor{white} & GWTC-5 & SIDM collapsed ($f_{\rm coll}$ marginalized) & $\gSISfm$ & $\PCloggSISfm$ \\
\midrule
\addlinespace[2pt]
\multirow{4}{*}{CIS} & Uninformative & Uninformative & $\CISuninfodds$ & --- \\
 & GWTC-5 & Uninformative & $\CISgwtcodds$ & --- \\
 & GWTC-5 & SIDM collapsed ($f_{\rm coll}=1$) & $\CISfone$ & $\PClogCISfone$ \\
\rowcolor{gray!15}[\tabcolsep][\tabcolsep] \cellcolor{white} & GWTC-5 & SIDM collapsed ($f_{\rm coll}$ marginalized) & $\CISfm$ & $\PClogCISfm$ \\
\midrule
\addlinespace[2pt]
\textit{NFW} & \textit{GWTC-5} & \textit{Collisionless CDM HMF} & \textit{$\mathit{\lesssim -6}$} & \textit{$\mathit{\lesssim -10}$} \\
\bottomrule
\end{tabular}
\caption{
Lensing posterior odds $\mathcal{O}^{\rm L}_{\rm unL}$ for GW231123 using the \texttt{NRSur7dq4} waveform model, and the frequentist probability $p^{\rm L}_{{\rm obs}\geq1}$ of detecting a GW231123-like lensed event from O1 to O4b.
For the PM lens (Sec.~\ref{subsec:PM_astro}), the $f_{\rm BH}$ and $f_{\rm PBH}$ rows are upper limits from observational constraints on the fraction of dark matter in general black holes (BHs) or primordial black holes (PBHs), assuming their mass is fine tuned to a constraint-satisfying optimal value $M_{\rm opt}$.
The last row assumes a mass function of intermediate-mass black hole (IMBH) lenses formed in star clusters.
For the gSIS and CIS lenses (Sec.~\ref{subsec:SIDM_astro}), the self-interacting dark matter (SIDM) rows assume a velocity dependent self-interacting cross section $\sigma(v)$ that collapses a fraction $f_{\rm coll}$ of halos gravothermally, either taking $f_{\rm coll} = 1$ or marginalizing over the $\sigma(v)$ parameters for a Yukawa potential.
Note that we assumed SIDM to be the correct DM model a priori for these rows.
The NFW entry (Sec.~\ref{subsec:NFW_astro}), corresponding to a collisionless cold dark matter (CDM) halo whose abundance follows the halo mass function (HMF), is an order-of-magnitude estimate obtained by rescaling the gSIS result to the NFW optical depth.
The last column $p^{\rm L}_{{\rm obs}\geq1}$ (Sec.~\ref{sec:rate}) is the frequentist probability of detecting at least one lensed event with parameters as atypical as GW231123 over O1--O4b, where dashes mark the rows with an uninformative lens prior for which it is not defined.
Error bars for $\mathcal{O}^{\rm L}_{\rm unL}$ denote the $90\%$ interval combining errors in the evidence integrals and the GWTC-5 hyperparameter uncertainties, while the uncertainties on $p^{\rm L}_{{\rm obs}\geq1}$ are obtained by sampling from the posterior of GW231123.
The highlighted rows use the priors that most faithfully represent our prior belief.}
\label{tab:odds_NRSur}
\end{table*}

In 1919, Eddington observed the gravitational lensing of light from distant stars by the Sun during a solar eclipse~\cite{Dyson:1920cwa}, and the measured lensing effect followed the predictions of Einstein's general theory of relativity (GR).
A century later, in 2015, the LIGO detectors observed gravitational waves (GWs) for the first time, verifying another major prediction of relativity~\cite{LIGOScientific:2016aoc}.
The combination of these two phenomena in GR leads to yet another elegant prediction, that the propagation of GWs themselves can be affected by gravity and be lensed~\cite{Ohanian:1974ys}.
Lensed GWs have not been observed confidently, but such an observation will unlock immense scientific potential in astrophysics, cosmology and fundamental physics, see Refs.~\cite{Grespan:2023cpa,Smith:2025axx,Chen:2026qtu} and references therein.

Lensed GWs can be divided into different categories according to the physical setup~\cite{1992grle.book.....S}.
When the wavelength of the GWs is small compared to the gravitational length scale of the lens, the GWs can be considered as rays traveling on lines by working in the geometrical optics limit.
In this regime, we would detect multiple GWs with similar morphologies (but different amplitudes and phases in general, depending on image type) at different times, corresponding to GWs from the same source arriving at the detector after traveling through different paths around the lens.
As this is analogous to electromagnetic (EM) wave lensing, we reuse the same terminology and call these copies ``images''. 
In the geometrical optics limit, the lensed GWs are easier to model and analyze because they are just copies of the unlensed GWs with a scaled complex amplitude (or Hilbert transformed for type II images~\cite{Dai:2017huk,Ezquiaga:2020gdt}) that can be computed with Fermat's principle.

On the other hand, if the wavelength of the GWs is comparable or longer than the gravitational length scale of the lens, we will need to work in the full wave-optics regime and consider diffraction effects~\cite{Ulmer:1994ij,Takahashi:2003ix}, see Ref.~\cite{Leung:2023lmq} for a review.
The theoretical modeling of diffraction lensing has been well studied for decades, with multiple schemes developed for rapid and accurate evaluation of the lensing amplification factor~\cite{Ulmer:1994ij,Diego:2019lcd,Feldbrugge:2019fjs,Cheung:2020okf,Guo:2020eqw,Mishra:2021xzz,Tambalo:2022plm,Shan:2022xfx,Shan:2023qvd,Villarrubia-Rojo:2024xcj,Wu:2025rvy}, while end-to-end data analysis pipelines have also been developed~\cite{Lai:2018rto,Wright:2021cbn,Seo:2021psp,Yeung:2023mbs,Cheung:2024ugg,Villarrubia-Rojo:2024xcj,Chakraborty:2024mbr,Yeung:2024pir,Su:2025xry,Caldarola:2025oxr}.
The expected rates of these diffracted events have been studied~\cite{Choi:2021bkx,Guo:2022dre}, although the results inevitably have large scatter due to uncertainties in the abundance and profile of dark matter halos.
More broadly, GW lensing has been proposed as a probe of the abundance and density profiles of dark matter halos and their substructure~\cite{Tambalo:2022plm,Tambalo:2022wlm,Caliskan:2023zqm,Cheung:2024ugg,Singh:2025uvp,Vujeva:2025nwg,Barsode:2024wda,Li:2026dai}.

Since the first detection of gravitational waves a decade ago~\cite{LIGOScientific:2016aoc}, gravitational-wave lensing has become an observational science, and lensed events have been actively searched for in the LIGO-Virgo-KAGRA (LVK) data~\cite{Haris:2018vmn,Li:2019osa,McIsaac:2019use,Hannuksela:2019kle, Liu:2020par,Dai:2020tpj,Lo:2021nae,LIGOScientific:2021izm, Janquart:2022wxc,Ali:2022guz,LIGOScientific:2023bwz,Ezquiaga:2023xfe,Li:2023zdl}, including diffraction-lensed ones~\cite{Lai:2018rto,Kim:2022lex,Chakraborty:2025maj}, although no confident candidates have been identified.

On November 23, 2023, the LIGO Hanford and Livingston detectors jointly detected the GW231123 gravitational-wave event~\cite{LIGOScientific:2025rsn}. 
Assuming that the event is sourced by a binary black hole (BBH) merger without lensing, its inferred properties are exceptional (see Refs.~\cite{Mould:2026nle,Tenorio:2026dcc} for a discussion of how ``exceptionality'' can be quantified).
The source-frame masses of the progenitor black holes (BHs) were inferred to be $137^{+23}_{-18} M_\odot$ and $101^{+22}_{-50} M_\odot$, while the spins are $0.90^{+0.10}_{-0.19}$ and $0.80^{+0.20}_{-0.52}$, making this the most massive and most highly spinning BBH ever detected.
Moreover, the parameter estimation (PE) results applying different waveform models are biased with respect to each other, implying that either the event lies in an unconventional parameter range in which the waveforms have high systematic errors, or the data cannot be explained well by a conventional BBH waveform model with a Gaussian noise model.
This motivates testing an alternative hypothesis where the BBH was diffraction-lensed, which could have artificially increased the inferred masses due to lensing magnification and affected the inferred spins due to a change in the morphology in the waveform.
In fact, there is a degeneracy between diffraction-lensed GWs and unlensed but precessing ones, so a GW inferred to be highly precessing by an unlensed model would naturally be a candidate for diffraction-lensing analysis~\cite{Shan:2025jpt}.

In this work, we re-analyze GW231123 assuming that it is lensed by a spherically symmetric halo modeled as a point mass (PM), generalized singular isothermal sphere (gSIS), and cored isothermal sphere (CIS) lens.
We compare the results against the unlensed analysis, and we take extra care to quantify our prior belief in each hypothesis to ensure that, to the best of our ability, the posterior odds that we quote are faithful.
We further complement the posterior odds with a frequentist estimate of how probable it is to detect a lensed event with properties like GW231123, given the abundance of lenses.
This work is complementary to recent lensing analyses of GW231123~\cite{LIGOScientific:2025cwb,Chan:2025kyu,Goyal:2025eqo,Hu:2025lhv,Shan:2025dcd,Chakraborty:2025pxt,Wang:2026yjk,Barsode:2026bcs}.
To the best of our knowledge, this is the first work to include diffuse halo lenses in the lensing model and to include the astrophysical prior of the source and lens parameters in the quoted posterior odds.
We find that, for the priors that we consider most faithful, the posterior odds do not favor the lensing hypothesis, and that detecting a lensed event as atypical as GW231123 is in any case unlikely, so we conclude that GW231123 is unlikely to be diffraction-lensed by an isolated object.

\section{Lens models}

\subsection{How the lens model affects our analysis}

Lensing is a gravitational phenomenon that depends on the density profile of the lens.
In this work, we will consider lenses with a spherically symmetric density profile, in particular the PM, gSIS and CIS lenses.
These lens models have been studied in detail in the literature (See e.g. Refs.~\cite{Tambalo:2022plm,Tambalo:2022wlm}).
We will assume that these lenses exist in isolation without any external matter field or gravitational potential.

Our analysis depends on the lens model in two ways.
First, the distortions in the waveform due to lensing depend on the lens model. 
Such distortions are encoded in the frequency-dependent amplification factor $F(f)$, and the lensed waveform $h_{\rm L}$ is related to the unlensed waveform $h_{\rm unL}$ by
\begin{equation}\label{eq:Fh}
    h_{\rm L}(f) = F(f) h_{\rm unL} (f). 
\end{equation}
The fact that the effects of lensing can be factored out as a factor $F(f)$ multiplying the unlensed waveform can be understood by noting that, in the time domain, lensing affects the waveform the same way at every time $t$. This means the time-domain lensed waveform can be obtained by convolving the unlensed waveform with a lensing-dependent function, which gives Eq.~\eqref{eq:Fh} after Fourier transforming into the frequency domain.

The second way in which the lens model will affect our subsequent analysis is through the astrophysical prior we use for the lensing hypothesis.
The procedure for constructing such a prior and reweighting our Bayesian inference runs is explained in more detail in Sec.~\ref{sec:astro_prior}, but the main idea is that we quantify the abundance of lenses in each redshifted lens mass bin, $dn/dM_{Lz}$, from which the optical depth and hence the prior probability of lensing follow.
While the natural mass scale in the lensing computation is the redshifted lens mass $M_{Lz}$, the abundance of lenses is more naturally specified in terms of another mass, such as the virial mass $M_{\rm vir}$ for a halo, and the conversion between the two depends on the lens model.

\subsection{Definition of length and mass scales}\label{subsec:length_scale}

In this work, we will assume that the lens is spherically symmetric, so that we can work on a 1D slice of the sky instead of the full 2D plane without loss of generality.
Let $D_L$, $D_S$ and $D_{LS}$ be the angular diameter distances from the observer to the lens, to the source, and between the lens and the source respectively.
Draw a central axis that extends from the observer to the center of the lens and passes through the source plane.
Define $\eta$ to be the perpendicular distance of the source from this axis on the source plane.
In the geometrical optics limit, i.e.\ assuming that the source emits rays, for a ray that reaches the observer, the ray would pass through the lens plane at a perpendicular distance $\xi$.
It is useful to define a length scale $\xi_0$ such that we can work with dimensionless quantities
\begin{align}
    x &= \dfrac{\xi}{\xi_0}, \label{eq:x_def}\\
    y &= \dfrac{\eta}{\eta_0},
\end{align}
where $\eta_0 = \xi_0 D_S/D_L$.
Without loss of generality, we can always define $y$ to be positive, and a negative $x$ means that the ray passes through the lens plane on the other side of the central axis.
We define the projected surface density of the lens to be
\begin{equation}
    \Sigma(\xi) = \int dz \, \rho(\xi, z),
\end{equation}
where $z$ is the line-of-sight direction perpendicular to the lens plane and $\rho$ is the 3D mass density of the lens.
We also define the convergence
\begin{equation}
    \kappa(x) = \dfrac{\Sigma(\xi_0 x)}{\Sigma_{\rm crit}},
\end{equation}
where $\Sigma_{\rm crit} = c^2 D_S/4\pi G D_LD_{LS}$.
Given a fixed $y$, for a ray that is able to reach the observer, it would have passed through the lens plane at location $x$ given implicitly by the lens equation,
\begin{equation} \label{eq:lens_eq}
    y = x(1 - \bar{\kappa}(x)),
\end{equation}
which depends on the mean convergence $\bar{\kappa}$ within a disk of dimensionless radius $|x|$ on the lens plane,
\begin{equation}
          \bar{\kappa} \equiv \dfrac{2}{x^2}\int^{|x|}_0 \kappa(x^\prime)x^\prime dx^\prime.
\end{equation}
The fact that the Eq.~\eqref{eq:lens_eq} depends on $\bar{\kappa}$ and not $\kappa$ can be derived from the shell theorem.
Note that in the physical setup, $y$ is the parameter that is free to change (i.e.\ where we place the source) and $x$ depends on $y$ via Eq.~\eqref{eq:lens_eq}. 
For a source at $y = 0$, if $\bar{\kappa}(x) = 1$ for all $x$, rays passing through any $x$ would be able to reach the observer.
In other words, $\Sigma_{\rm crit}$ is defined such that if the lens has an infinite 2D extent on the lens plane with a uniform surface density of $\Sigma_{\rm crit}$, all rays emitted by the source will be focused to the same point at the observer. 

All lensing observables, e.g. the magnification and time delay of images in the geometrical optics limit, are derived from Eq.~\eqref{eq:lens_eq}.
Therefore, $\bar{\kappa}(x)$ controls the significance of all lensing effects. 
If $\bar{\kappa}(x) \ll 1$, $y \approx x$, which is the trivial setup where lensing effects are negligible.
If $\bar{\kappa}(x) \sim 1$ then the fractional difference between $y$ and $x$ is $O(1)$, i.e.\ the ray is significantly deflected.
Therefore, the strength of lensing depends on how $\bar{\kappa}(x)$ compares to $1$.
For most astrophysically realistic lens models, $\bar{\kappa}(x)$ decreases monotonically with $x$, so if $\kappa(0) > 1$ there is an $x_E$ such that $\bar{\kappa}(x_E) = 1$.
Defining
\begin{equation}
    x_E \equiv \dfrac{\xi_E}{\xi_0},
\end{equation}
we call $\xi_E$ the Einstein radius.
If we adjust the length scale used in Eq.~\eqref{eq:x_def} such that $\xi_0 \equiv \xi_E$, we get $x_E \equiv 1$.
With this choice of scale, to have negligible lensing effects, i.e.\ $\bar{\kappa}(x) \ll 1$, we must be in the $x \gg x_E \equiv 1$ regime, so $y \gg 1$.
To have significant lensing effects, i.e.\ $\bar{\kappa}(x) \sim O(1)$, we must have $x \sim O(1)$ and $y \lesssim O(1)$.
In other words, by choosing the length scale to be $\xi_E$, it is easy to gauge whether lensing effects will be significant just by looking at how $y$ compares with $1$.

The length scale $\xi_0$ can be converted to a mass scale $M_{Lz}$,
\begin{equation}\label{eq:MLz_def}
    M_{Lz} \equiv \dfrac{c^2 \xi_0^2}{4 G d_{\rm eff}},
\end{equation}
with
\begin{equation}\label{eq:deff_def}
    d_{\rm eff} \equiv \dfrac{D_L D_{LS}}{(1 + z_L) D_S}.
\end{equation}
We call $M_{Lz}$ the redshifted lens mass.
This is the redshifted projected mass contained within a disk of radius $\xi_0$ on the lens plane.
In principle, $M_{Lz}$ is a mass scale as arbitrary as the definition of the length scale $\xi_0$, and fixing one fixes the other.
For the choice $\xi_0 \equiv \xi_E$, the mass scale $M_{Lz}$ is the redshifted Einstein mass, which is intuitively the mass of the part of the lens responsible for the lensing effects.

When analyzing the same data with different lens models, if we choose $\xi_0 \equiv \xi_E$ respectively for each model, their inferred $M_{Lz}$ and $y$ should be comparable to each other.
The choice of $\xi_0$ below for the PM and gSIS models will be exactly $\xi_E$, while that for CIS will be $\xi_0 = \xi_E/\sqrt{1 - 2 x_c}$ following the standard convention in the literature, which is approximately $\xi_E$ for small $x_c$.
Note that for an NFW profile lens, the usual convention is $\xi_0 = r_s$, where $r_s$ is the scale radius of the profile, which can be many orders of magnitude larger than $\xi_E$.

\subsection{Lens-model specific quantities}\label{subsec:lens_models}

\subsubsection{Point-mass lens (PM)}

As mentioned in Sec.~\ref{subsec:length_scale}, for the PM lens, we define $M_{Lz}$ to be the redshifted mass of the point mass, leading to the trivial relation
\begin{equation}
    M_{Lz}^{\rm PM} \equiv (1 + z_L)M^{\rm PM},
\end{equation}
where the superscript PM reminds us that the definition of lens mass is lens-model-dependent.
The point mass in question is equivalent to a Schwarzschild black hole, with $M^{\rm PM}$ simply corresponding to its mass.
Then, $\xi_0$ can be obtained with Eq.~\eqref{eq:MLz_def} and found to be the Einstein radius $\xi_E$.
The only parameter fixing the form of the lens is $M_{Lz}$.

The lensing amplification factor $F$ that appears in Eq.~\eqref{eq:Fh} for the PM lens can be obtained analytically~\cite{Nakamura:1997sw},
\begin{equation}
    F^{\rm PM}(w) = u^{iu} e^{(\pi/2 - 2it_{\rm min})u} \Gamma(1 - i u) {}_1F_1(iu,1;iuy^2)
\end{equation}
where $w \equiv 8\pi G M_{Lz} f/c^3$ is the dimensionless frequency, $\Gamma$ is the gamma function, $_1F_1$ is the Kummer confluent hypergeometric function, and we have adopted the notation of Ref.~\cite{Villarrubia-Rojo:2024xcj} with $u = w/2$, $t_{\rm min} = (x_{\rm min} - y)^2/2 - \ln(x_{\rm min})$ and $x_{\rm min} = \left(y + \sqrt{y^2 + 4}\right)/2$.
As mentioned in Sec.~V.A of Ref.~\cite{Villarrubia-Rojo:2024xcj}, this can be evaluated rapidly by using different asymptotic formulas and series expansions in different limits, as is implemented in the \texttt{GLoW} package developed by the same authors.
To integrate with the \texttt{glworia} package better, we reimplement the \texttt{GLoW} point-mass lens computation in \texttt{python} within the \texttt{glworia} package.

\subsubsection{Generalized singular isothermal sphere lens (gSIS)}

The density profile of the gSIS lens, also called the sphere power-law (SPL) lens, is 
\begin{equation} \label{eq:gSIS_profile}
    \rho(r) = \rho_0\left( \dfrac{r_0}{r}\right)^{k + 1}.
\end{equation}
The length scale $\xi_0$ is chosen to be
\begin{equation} \label{eq:xi0_gSIS}
    \xi_0 = \left(\dfrac{2 \beta_k}{2 - k} \dfrac{\rho_0 r_0^{k+1}}{\Sigma_{\rm crit}}\right)^{1/k},
\end{equation} 
where $\Sigma_{\rm crit} = c^2 D_S/4\pi G D_L D_{LS}$ and $\beta_k = \sqrt{\pi}\Gamma(k/2)/\Gamma((k+1)/2)$.
This choice of $\xi_0$ fixes the definition of $M^{\rm gSIS}_{Lz}$ via Eq.~\eqref{eq:MLz_def}, and can be shown to be equivalent to $\xi_E$. 
Note that while Eq.~\eqref{eq:gSIS_profile} seems to depend on three parameters $\rho_0$, $r_0$ and $k$, only the combination $\rho_0 r_0^{k + 1}$ enters in both Eqs.~\eqref{eq:gSIS_profile} and~\eqref{eq:xi0_gSIS}, so once the two degrees of freedom $M_{Lz}$ (or $\xi_0$) and $k$ are fixed, the profile is determined completely.
The singular isothermal sphere (SIS) lens is recovered by setting $k = 1$. 
The density profile for gSIS is steeper than SIS for $k > 1$, making it more compact and the lensing effects stronger.
The opposite is true when $k < 1$.

To the best of our knowledge, there is no closed-form expression for $F^{\rm gSIS}(w)$, so we have to compute it numerically.
This requires solving a 2D Kirchhoff diffraction integral numerically.
We use the implementation in the \texttt{glworia} package, which evaluates the integral in the time domain by performing contour integrals around contours of different time delays on the image plane, a method pioneered by Ref.~\cite{Ulmer:1994ij}.
The numerical integration is not rapid enough for Bayesian inference, which often requires $\sim 10^7$ likelihood evaluations to sample the posterior distribution adequately, so we compute the integral on a grid over the lensing parameter space and interpolate in between, see Ref.~\cite{Cheung:2024ugg} for more details.

\subsubsection{Cored isothermal sphere lens (CIS)}

The density profile of the CIS lens, also called the non-singular isothermal sphere (NIS), is 
\begin{equation}\label{eq:CIS_profile}
    \rho(r) = \rho_0 \dfrac{r_c^2}{r^2 + r_c^2},
\end{equation}
and we choose $\xi_0$ to be
\begin{equation}
    \xi_0 = \dfrac{2 \pi \rho_0 r_c^2}{\Sigma_{\rm crit}},
\end{equation}
which fixes the definition of $M_{Lz}$ through Eq.~\eqref{eq:MLz_def}.
The length scale $r_c$ is the scale of a central core that prevents $\rho(r)$ from diverging at $r = 0$.
We define the dimensionless core radius 
\begin{equation}
    x_c \equiv \dfrac{r_c}{\xi_0}.
\label{eq:dim_core_radius}
\end{equation}
The form of the CIS lens is fixed by the two parameters $M_{Lz}$ and $x_c$.
The SIS lens is recovered for $x_c = 0$.
The length scale used can be shown to satisfy $\xi_0 \equiv \xi_E/\sqrt{1 - 2 x_c}$, so $\xi_0 \approx \xi_E$ for $x_c \ll 0.5$.

Similar to the gSIS case, to the best of our knowledge, there is no closed-form expression for $F^{\rm CIS}(w)$, so we compute the amplification factor numerically with the implementation in \texttt{glworia}.

\section{Bayesian inference with uninformative priors}
\label{sec:uninformative}

\subsection{Bayesian inference code for diffraction-lensed GWs}

To determine whether GW231123 is lensed, we perform Bayesian parameter inference on the detector data.
This requires calling the amplification factor $F$ and the unlensed waveform $h_{\rm unL}$ for computing the likelihood, as well as an algorithm for sampling the high-dimensional posterior distribution.
We use the \texttt{glworia} package to perform the inference, which constructs an interpolation table for $F$ if necessary, and interfaces with the \texttt{bilby}~\cite{bilby_paper, bilby_doi} package for calling $h_{\rm unL}$ and performing nested sampling with \texttt{dynesty}~\cite{Speagle:2019ivv,sergey_koposov_2025_17268284}.

As mentioned in Sec.~\ref{subsec:lens_models}, we will make use of the PM, gSIS and CIS lens models in this work.
The $F^{\rm PM}(w)$ has an analytic form that depends on $M_{Lz}$ (when converting the frequency from $w$ to $f$) and $y$, so we compute it in real time for likelihood evaluation using the asymptotic expressions and expansions listed in Ref.~\cite{Villarrubia-Rojo:2024xcj}.
On the other hand, in addition to $M_{Lz}$ and $y$, $F^{\rm gSIS}$ depends also on the parameter $k$ and $F^{\rm CIS}$ depends on $x_c$, and their computation requires numerically computing a Kirchhoff integral, so we make use of an interpolator.
The interpolation scheme breaks down when the source is near a caustic of the lens, so we will excise those regions of parameter space in our analysis by setting the prior probability density to zero.
Note that analytic results exist at the caustic~\cite{Ezquiaga:2025gkd}, but these have not been implemented in \texttt{glworia}.
The details of the computation of the integral, the interpolation scheme and other aspects of the \texttt{glworia} code are explained in Ref.~\cite{Cheung:2024ugg}.

In this work, we focus on the strong lensing regime $0.01 < y < 2.0$. 
We also apply the bounds $0.1 < k <1.9$ for the gSIS model and $0 < x_c<0.5$ for the CIS model.

\subsection{Parameter estimation results}

\begin{figure*}
    \centering
    \includegraphics[width=0.45\linewidth]{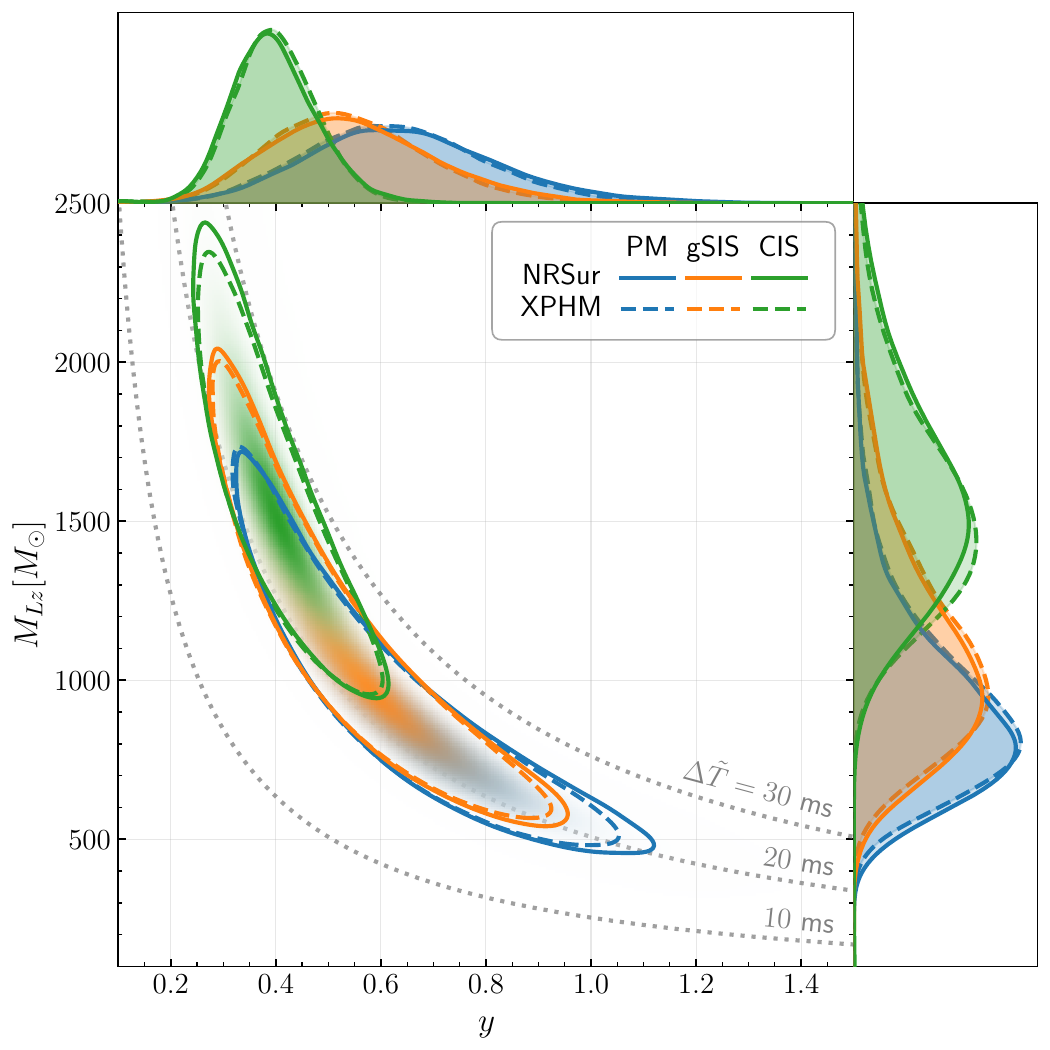}\\
    \includegraphics[width=0.45\linewidth]{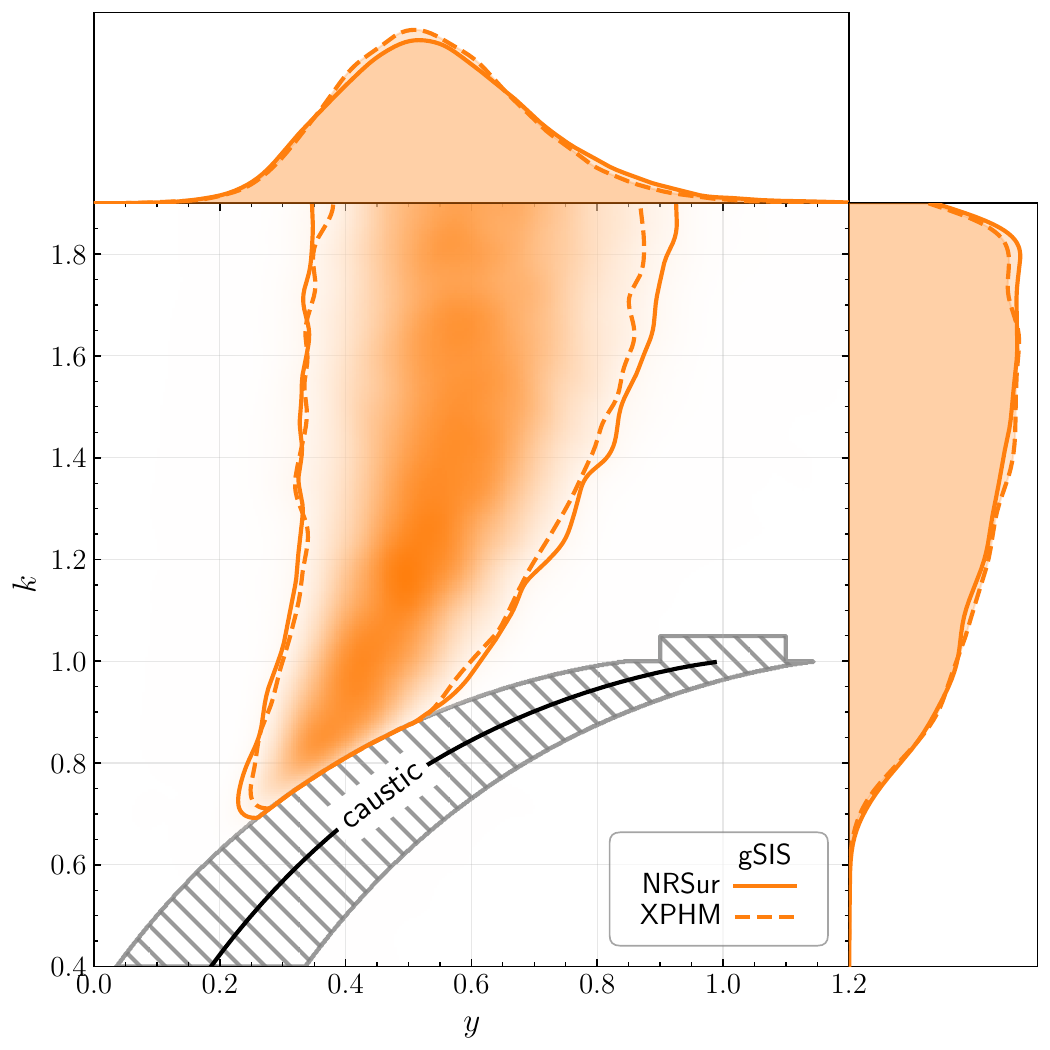}
    \includegraphics[width=0.45\linewidth]{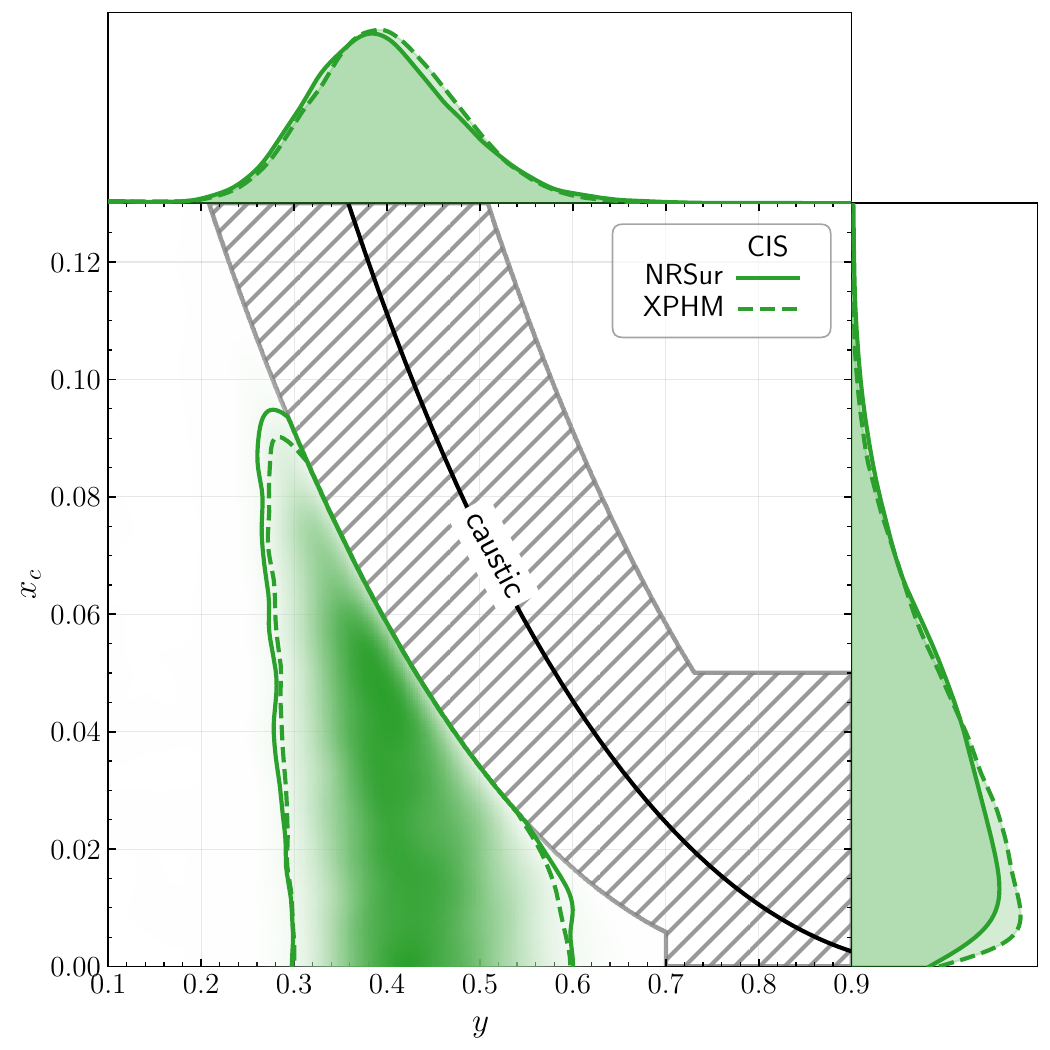}
    \caption{The posterior probability distribution of the lens parameters when the point mass (PM, blue), generalized singular isothermal sphere (gSIS, orange) and cored isothermal sphere (CIS, green) models are applied to GW231123, using the \texttt{NRSur7dq4} (NRSur, solid contours) and \texttt{IMRPhenomXPHM-SpinTaylor} (XPHM, dashed contours) waveform models to model the source gravitational waveform.
    \textit{Top:} the posterior in the redshifted lens mass ($M_{Lz}$) vs dimensionless impact parameter ($y$) plane.
    The shape of the posterior approximately follows lines of constant effective time delay between the two dominant images $\Delta \tilde{T} = 8 M_{Lz} y$ (gray dotted lines).
    \textit{Bottom:} The posterior distribution of the additional lens parameters: density profile power-law slope $k$ (see Eq.~\eqref{eq:gSIS_profile}) for the gSIS model (left), and the dimensionless core radius $x_c$ (see Eq.~\eqref{eq:dim_core_radius}) for the CIS model (right) vs $y$.
    We excise the region near the caustic curves (gray hatches) by setting the prior probability to zero because our lensing amplification factor computation is not accurate in those regimes.}
    \label{fig:lens_params_posterior}
\end{figure*}

\begin{figure*}
    \centering
    \includegraphics[width=0.45\linewidth]{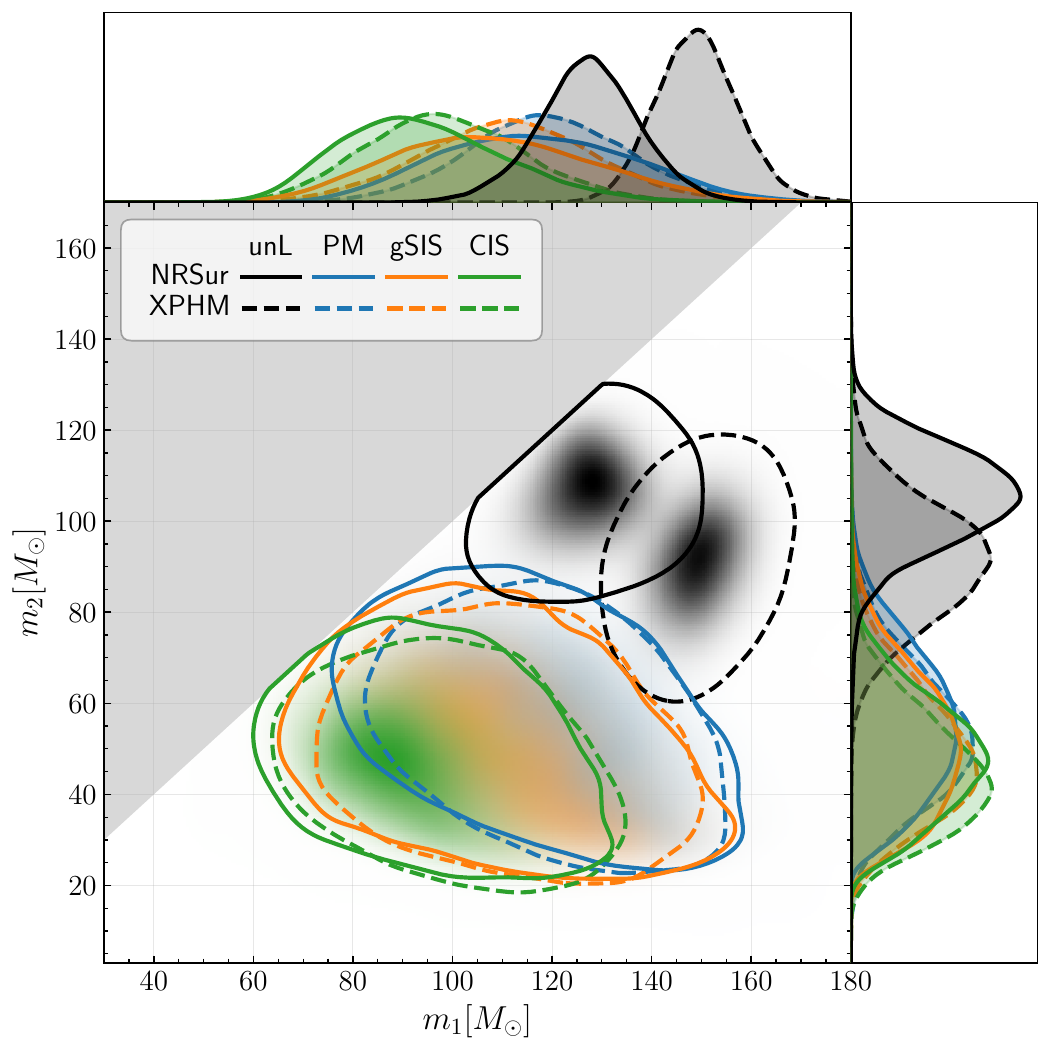}
    \includegraphics[width=0.45\linewidth]{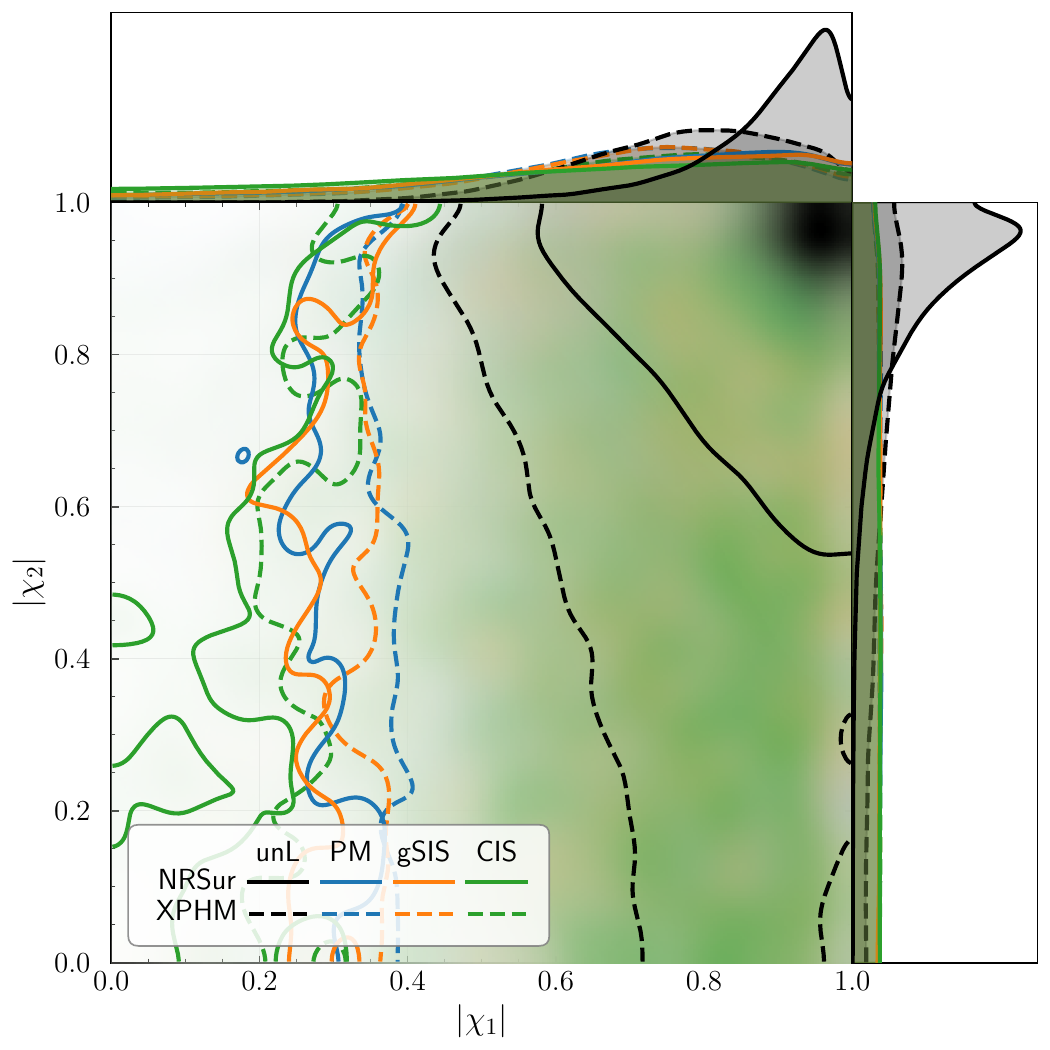}
    \caption{The posterior probability distribution of the source BBH parameters when the unlensed (unL, black), PM (blue), gSIS (orange) and CIS (green) models are applied to GW231123, using the \texttt{NRSur7dq4} (NRSur, solid contours) and \texttt{IMRPhenomXPHM-SpinTaylor} (XPHM, dashed contours) waveform models to model the source gravitational waveform.
    \textit{Left:} the posterior distribution of the source-frame primary mass $m_1$ and secondary mass $m_2$.
    The gray region ($m_1 < m_2$) is forbidden by the convention that $m_1$ is the more massive component.
    \textit{Right:} the posterior distribution of the magnitude of the spin of the primary $|\chi_1|$ and secondary $|\chi_2|$ black hole.
    The posteriors shift to lower masses and lower spins and the two waveform models give compatible results when a lens model is applied.}
    \label{fig:source_params_posterior}
\end{figure*}

\begin{figure*}
    \centering
    \includegraphics[width=0.9\linewidth]{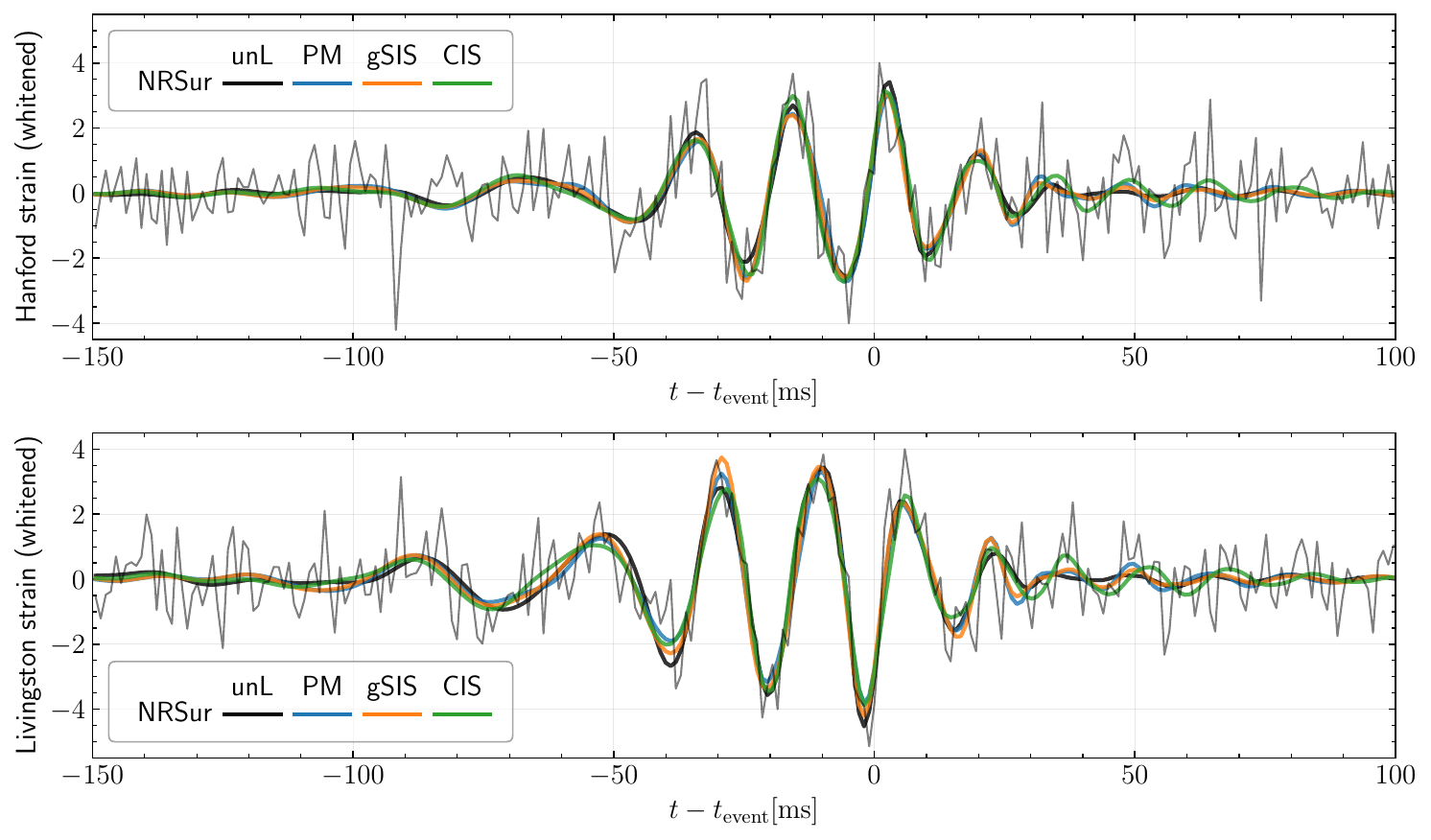}
    \caption{The maximum likelihood whitened time-domain waveform of the unlensed (unL, black), PM (blue), gSIS (orange) and CIS (green) analyses using the \texttt{NRSur7dq4} (NRSur) waveform model superimposed on the whitened detector data strain (gray) of the LIGO Hanford (top) and Livingston (bottom) detectors for GW231123.
    The lensed and unlensed waveforms deviate visually at around $t - t_{\rm event} \gtrsim 20 \, \rm ms$, consistent with the rest of the literature studying this event.
    We do not show the \texttt{IMRPhenomXPHM-SpinTaylor} waveforms to avoid clutter.}
    \label{fig:waveform}
\end{figure*}

As hinted in the title of this work, the prior distribution is important when it comes to determining whether GW231123 is lensed. 
Nonetheless, we will first perform PE with uninformative priors, i.e.\ simple, commonly used priors that are not informed by astrophysics.
The posterior samples can be reweighted to new priors later if appropriate.

The settings of our PE run mostly follow those used in the LVK analysis of GW231123.
We use 8 seconds of data from the LIGO Hanford and Livingston detectors, starting from 6 seconds before the trigger time of the event.
We also perform an unlensed run that serves as a baseline for comparison.
The uninformative prior and reparameterization choices that we use to facilitate efficient sampling are described in detail in Appendix~\ref{app:priors}.
In particular, we sample an effective time-delay parameter $\Delta\tilde{T} = 8 M_{Lz} y$, which is approximately the time delay between the first (type I) and second (type II) image in the regime $y \lesssim 1$ for all the lens models considered ($\Delta\tilde{T}$ is exactly the time delay for the SIS lens). We also sample $\tilde{d}_L = d_L \, y$, which approximately breaks the degeneracy between $d_L$ and the magnification of the waveform (the magnification being approximately inversely proportional to $y$).

We find that the posteriors for the lensed parameters are well constrained for all three of the lensed models, and that both waveform models used, i.e.\ \texttt{NRSur7dq4}~\cite{Varma:2019csw} and \texttt{IMRPhenomXPHM-SpinTaylor}~\cite{Pratten:2020ceb,Colleoni:2024knd} give compatible posterior distributions.
As shown in the top panel of Fig.~\ref{fig:lens_params_posterior}, $M_{Lz}$ and $y$ are measured meaningfully, with posteriors constrained well compared to the prior distributions for the parameters.
None of the runs returned posteriors that rail against the upper bound of $y = 2.0$, which would mean that lensing effects are unmeasurable.
The inferred $M_{Lz}$ and $y$ are centered at different values for each lens model, which is expected because the lensing effects are different across lens models and because of the model-dependent definition of $\xi_0$, which affects $M_{Lz}$ and $y$.
Nonetheless, all three lens models give posteriors lying approximately on a constant line of $\Delta\tilde{T} \sim 22 \, {\rm ms}$, indicating that all models agree on the inferred time delay between the two dominant images.
Such an agreement is expected because it is a model-independent feature of the data: it implies that the data can be fitted well by superimposing two unlensed waveforms with similar morphologies offset in time by $\sim 22 \, {\rm ms}$, which is essentially what a lensed waveform is if we neglect diffraction effects, see Appendix~\ref{app:insights} for more details.
This number is consistent with the one quoted in recent analyses of GW231123 (e.g. Refs.~\cite{Goyal:2025eqo, Hu:2025lhv, Chan:2025kyu}).

For the diffuse halo lens models gSIS and CIS, there is one additional lens parameter compared to the PM case. 
In the bottom two panels of Fig.~\ref{fig:lens_params_posterior}, we show that the additional lens parameter is well constrained in both cases.
The power-law index $k$ of the gSIS model rails against the upper bound of $k = 1.9$, while the dimensionless core size $x_c$ rails against the lower bound of $x_c = 0$, both implying that the data favor more concentrated halos.
Note that the SIS lens is recovered for $k = 1.0$ in the gSIS model or $x_c = 0$ in the CIS model, and the posteriors in these two slices agree as expected.

Both posteriors rail against the excised region close to the caustic curve on the $k$-$y$ or $x_c$-$y$ planes.
This implies that the excision has inevitably affected the posteriors.
However, as crossing the caustic entails a ``phase transition'' in which the number of stationary points in the time-delay surface (corresponding to the number of images in the geometrical optics approximation) changes, the amplification factor $F$ changes qualitatively.
Even if we were able to compute $F$ accurately close to the caustic and do not have to excise the region, it is expected that the posterior will lie on one side of the caustic if the lensing effects are well measured.
Therefore, we do not expect the excision to have affected our results qualitatively.
Note that for the gSIS model, the number of images also changes when crossing the $k = 1$ line (SIS limit)~\cite{Tambalo:2022wlm}, but the behavior of the amplification factor is smooth, so no pathologies are introduced.

\begin{figure}
    \centering
    \includegraphics[width=0.99\linewidth]{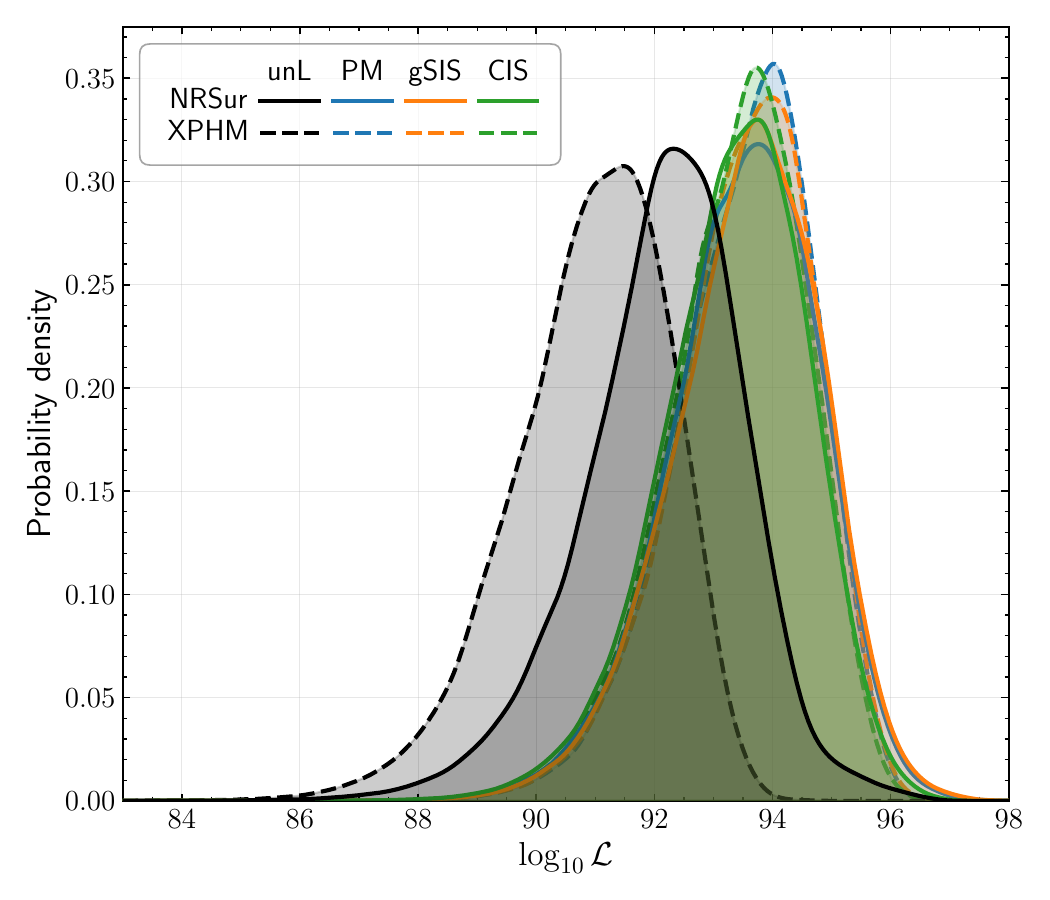}
    \caption{The distribution of likelihood for the posterior samples of each lens and waveform model.
    The bulk of the distribution for the lensed models agree with each other and are shifted higher compared to the unlensed ones.}
    \label{fig:likelihood}
\end{figure}

As mentioned in Sec.~\ref{sec:intro}, the lensing interpretation of GW231123 depends on our prior belief about the rarity of seeing an unlensed GW231123-like event versus that of seeing a lensed one.
This would depend on the inferred properties of the source BBH corresponding to the lensed models versus the unlensed one, because BBHs with high mass and high spin progenitors are expected to be rarer.
As shown in Fig.~\ref{fig:source_params_posterior}, defining the primary BH as the more massive BH in the source BBH, the inferred primary mass $m_1$, secondary mass $m_2$ and magnitudes of the spins of the primary $|\chi_1|$ and secondary $|\chi_2|$ all shift lower for the lensed analyses compared to the unlensed one.
While the two waveform models give posteriors with different peak locations or shapes for the unlensed model, the posteriors of the lensed models are similar across waveform models.

Other than the shape of the posteriors, it is also instructive to look at the goodness of fit to the data.
In Fig.~\ref{fig:waveform}, we show the inferred maximum likelihood waveform for all of the models employed using the \texttt{NRSur7dq4} waveform model. 
All models fit the data well by eye.
In fact, as shown in Fig.~\ref{fig:likelihood}, the lensed models give posterior samples with a higher likelihood value $\Delta \log_{10} \mathcal{L} \sim O(1)$ compared to the unlensed one.
This could be attributed to a better fit of the data in the ringdown stage where the lensed waveform has a slower decay.
Such a behavior in the lensed waveform can be understood by noting that if we neglect diffraction, in the geometrical optics limit the waveform is approximately a superposition of two waveforms with a time delay of $\sim 22 \, {\rm ms}$, meaning that the decay in the waveform is delayed by approximately the same amount of time.

All the results above seem to suggest that GW231123 is a promising candidate for a diffraction-lensed GW event:
the lensed parameters are well constrained; 
the source parameters shift to lower masses and spins, which are more common astrophysically;
the lensed waveforms fit the data better.
However, in the above analysis we have not included the fact that the lensed model has strictly more parameters than the unlensed one, and that the probability of lensing is inherently low.
Also, while the masses and spins of the BBH are inferred to be lower for the lensed case, the masses still overlap with the pair-instability supernova (PISN) mass gap~\cite{Woosley:2016hmi,Farmer:2019jed,Woosley:2021xba,Tong:2025wpz}, and the spin of the primary $|\chi_1|$ is still high.

To better quantify the support of the lensing hypothesis, we have to compute the posterior odds that account for the astrophysical prior in the source and lens parameters, as well as the probability of lensing.
We have deliberately avoided presenting a Bayes factor in the preceding discussion.
This is because a Bayes factor that does not account for the astrophysical priors could be misleading, especially given that the source parameters inferred with an unlensed model are outliers in the population of observed BBHs, and that lensing effects are expected to be rare due to a low optical depth.

\section{Reweighting to astrophysical priors}\label{sec:astro_prior}

\subsection{Priors and posterior odds}

In this section, we compute the posterior odds of the lensed versus unlensed hypotheses.
We reweight the Bayesian evidence of our nested sampling runs by transforming from the uninformative priors to astrophysically informed priors.
For the unlensed model, the parameters of the model can be represented by the parameter space vector $\vec{\theta}_{\rm S}$.
We label this vector with the subscript $\rm S$ (for source) to remind ourselves that these are parameters related to the properties of the source BBH and have nothing to do with lensing.
Note that $\vec{\theta}_{\rm S}$ includes both intrinsic parameters, like the masses and spins, and the extrinsic parameters, like the sky location and luminosity distance of the source.
For the lensed model, in addition to the same parameters $\vec{\theta}_{\rm S}$, there are also the lensed parameters $\vec{\theta}_{\rm L}$.

Before quantifying our astrophysical priors, we ought to address an important caveat directly related to the main conclusion of this work.
To perform model selection between two hypotheses $\mathcal{H}_0$ and $\mathcal{H}_1$ in Bayesian statistics, we compute the posterior odds $\mathcal{O}^1_0$,
\begin{equation}
   \mathcal{O}^1_0 \equiv \dfrac{P(\mathcal{H}_1|d)}{P(\mathcal{H}_0|d)} = \dfrac{P(d|\mathcal{H}_1)}{P(d|\mathcal{H}_0)}\dfrac{P(\mathcal{H}_1)}{P(\mathcal{H}_0)},
\end{equation}
where $P(d|\mathcal{H}_1)/P(d|\mathcal{H}_0)$ is the Bayes factor $\mathcal{B}^1_0$ and $P(\mathcal{H}_1)/P(\mathcal{H}_0)$ is the prior odds.
In our work, $\mathcal{H}_1 \equiv {\rm L}$ is the lensing hypothesis, and $\mathcal{H}_0 \equiv {\rm unL}$ is the unlensed hypothesis.
Both the lensed and unlensed hypotheses could be rare for GW231123, and there is no reason a priori to believe that the prior odds $P({\rm L})/P({\rm unL})$ are equal to one, so it cannot be neglected.

As we discuss below, defining the lensing hypothesis requires us to set bounds on the lens parameters, which changes the values of $\mathcal{B}$ and the prior odds while keeping $\mathcal{O}$ unchanged.
Therefore, rather than working with $\mathcal{B}$ and prior odds separately and multiplying them at the end, we find that it is conceptually simpler to compute $\mathcal{O}$ directly. 
We can absorb the prior odds into the prior distribution,
\begin{equation}
    \tilde{\pi}(\vec{\theta}|\mathcal{H}) = {\pi}(\vec{\theta}|\mathcal{H}) P(\mathcal{H}),
\end{equation}
where $\int {\pi}(\vec{\theta}|\mathcal{H}) d\vec{\theta} = 1$ by construction, implying
\begin{equation}\label{eq:prior_norm_convention}
    \int \tilde{\pi}(\vec{\theta}|\mathcal{H}) \, d\vec{\theta} = P(\mathcal{H}).
\end{equation}
The tilde on top of $\tilde{\pi}$ reminds us that it is not normalized within the prior bounds.
Then, the numerator and denominator of $\mathcal{O}$ can be computed directly,
\begin{equation}
    P(\mathcal{H}|d) = \dfrac{ \int \mathcal{L}(d |\vec{\theta}, \mathcal{H})\tilde{\pi}(\vec{\theta}|\mathcal{H})d\vec{\theta}}{P(d)},
\end{equation}
which is just the usual evidence integral but with an unnormalized prior, divided by $P(d)$ which cancels when taking the ratio between the two hypotheses.
In other words, $\mathcal{O}$ is equivalent to $\mathcal{B}$ with the prior $\pi$ replaced by $\tilde{\pi}$.

For the unlensed and lensed models, we can write the astrophysical prior as
\begin{align}
    \tilde{\pi}(\vec{\theta}_{\rm S} | {\rm unL}) &= \tilde{\pi} (\vec{\theta}_{\rm S}), \\
    \tilde{\pi}( \vec{\theta}_{\rm L}, \vec{\theta}_{\rm S}| {\rm L}) &= \tilde{\pi}(\vec{\theta}_{\rm L} | z_S)\tilde{\pi} (\vec{\theta}_{\rm S}), \label{eq:pi_lens}
\end{align}
where ${\rm unL}$ and ${\rm L}$ denote the unlensed and lensed hypotheses, and on the second line we have assumed that the prior on $\vec{\theta}_{\rm L}$ does not depend on any parameter in $\vec{\theta}_{\rm S}$ except for the source redshift $z_S$.

This convention has different practical implications for the source and lens parameters.
For the source parameters, the same astrophysical prior $\tilde{\pi}(\vec{\theta}_{\rm S})$ enters the evidence integral for both the lensed and unlensed hypotheses.
Any overall normalization factor therefore cancels exactly in the evidence ratio, provided the same prior bounds are used for both models.
We do apply bounds to the source parameters. 
For example, we use a lower bound of $\mathcal{M} > 40 \, \rm M_\odot$ on the detector-frame chirp mass during PE.
The GWTC-5 astrophysical prior carries significant weight below this bound, so strictly speaking the prior does not integrate to one within our PE bounds.
This is not a concern. 
Since the same truncation is applied in both models, the associated normalization factor cancels in the ratio, and the astrophysical prior affects only the shape of the integrand within the bounds, which is the desired outcome.

The lens parameters $\vec{\theta}_{\rm L}$ appear only in the lensed model and have no counterpart in the unlensed evidence integral.
Their normalization therefore does not cancel in the evidence ratio, and the integral $\sim \int \tilde{\pi}(\vec{\theta}_{\rm L} | z_S) \, d\vec{\theta}_{\rm L}$ directly contributes to the quoted posterior odds.

We define the lensing hypothesis as $y < y_{\rm cr}$, i.e.\ $y_{\rm cr}$ is the critical $y$ within which we define the wave to be lensed.
In this work, we will use $y_{\rm cr} \equiv 2$ as lensing effects with $y \gtrsim 2$ are generally negligible at current sensitivity and are already well described by the unlensed model, and, since gravity is long-range, any event is formally lensed to some degree, so a finite cutoff is necessary for a meaningful dividing line between the two hypotheses.
Under this definition, $P({\rm L})$ equals the fraction of GW events with $y < 2$, i.e.\ the optical depth (cf. Eq.~\eqref{eq:opt_depth} below).
We also impose a lower bound $y > 0.01$ to avoid numerical difficulties near the caustic at $y = 0$, but this removes only a negligible part of the prior volume.

If we normalized $\tilde{\pi}(\vec{\theta}_{\rm L})$ to one within the PE bounds $0.01 < y < 2$, we would be asserting $P({\rm L}) = 1$, effectively ignoring the prior odds entirely and using the Bayes factor in place of the posterior odds.
This would artificially inflate the support of the lensed hypothesis.
Instead, we compute $\tilde{\pi}(\vec{\theta}_{\rm L})$ from the astrophysical lensing rate and leave it unnormalized so that its integral within the PE bounds equals the optical depth.
The detailed calculation is given in Sec.~\ref{subsec:lens_astro}.

In summary, while the shape of the source parameter prior is important, its normalization cancels in the evidence ratio and is therefore irrelevant.
On the other hand, both the shape and normalization of the lens parameter prior are important: its normalization encodes the prior probability of lensing and must be computed carefully from the optical depth.

\begin{figure*}
    \centering
    \includegraphics[width=0.99\linewidth]{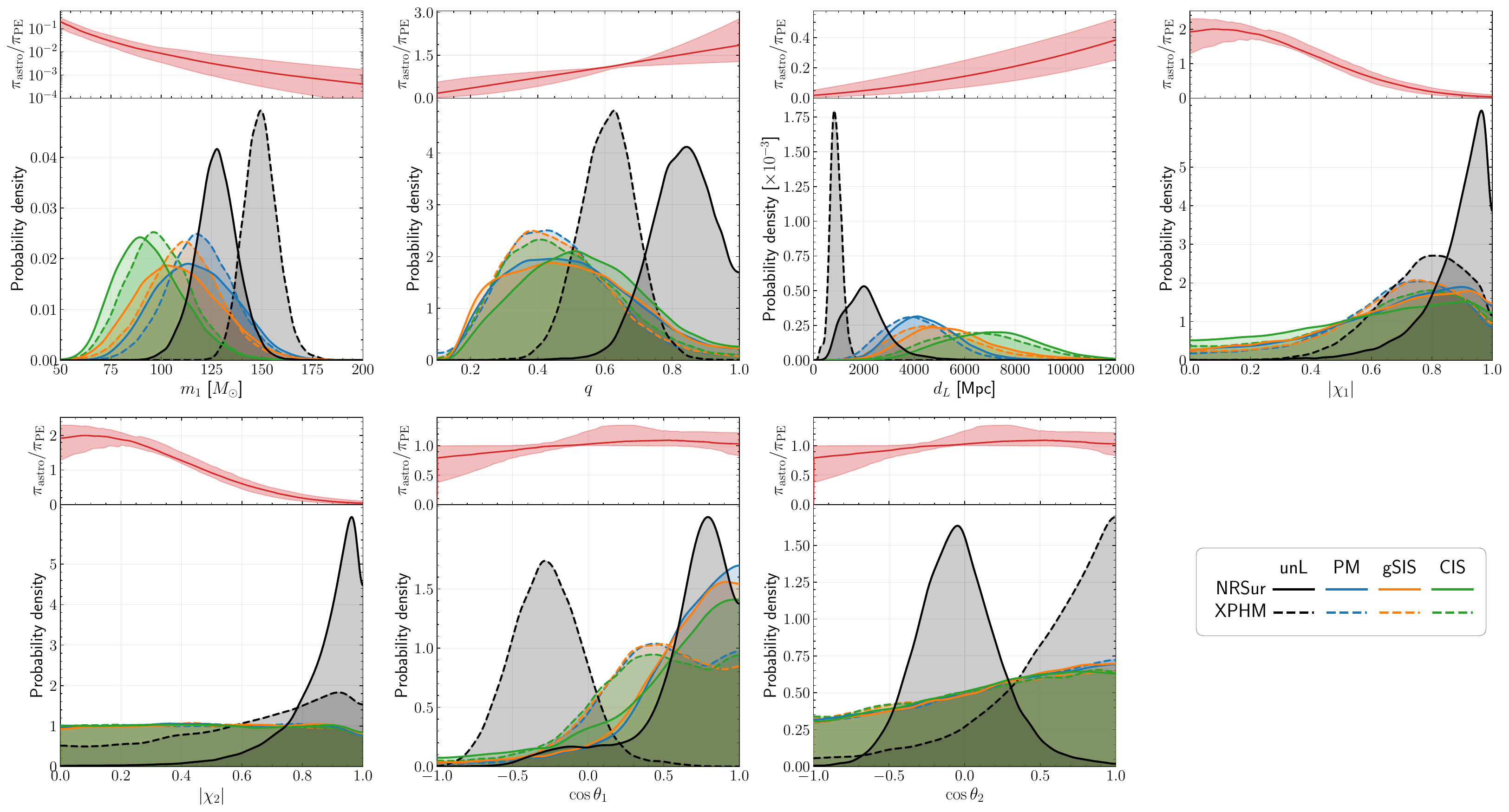}
    \caption{\textit{Top panels: } the prior reweighting factor $\pi_{\rm astro}/\pi_{\rm PE}$ (red curve), which is the ratio between the astrophysical prior (inferred from population of events in GWTC-5) and the uninformative prior used in PE. \textit{Bottom panels:} the posterior distribution for each lens and waveform model under the uninformative PE prior for the source-frame primary mass $m_1$, mass ratio $q$, luminosity distance $d_L$, spin magnitudes of the primary $|\chi_1|$ and secondary $|\chi_2|$, and spin tilt angles of the primary $\cos \theta_1$ and secondary $\cos \theta_2$.
    The reweighting factor favors the lensed hypothesis for $m_1$, $d_L$, $|\chi_1|$ and $|\chi_2|$, and favors the unlensed hypothesis for $q$. 
    }
    \label{fig:astro_prior}
\end{figure*}

\subsection{Source parameter astrophysical prior}\label{subsec:source_astro}

To quantify the astrophysical prior probability of the source parameters, we use the distribution of parameters inferred from the population of events in GWTC-5~\cite{LIGOScientific:2026ctl}, which includes all BBHs detected confidently by pipelines within the LVK Collaboration up to the second part of the fourth observing run (O4b).
The population includes GW231123 itself with the assumption that it was unlensed; using such an astrophysical prior would double count the event and inflate the support for the unlensed hypothesis.
Therefore, in principle, one should leave out the event in question when inferring an astrophysical prior for said event from a population to avoid double counting.
However, we choose to use the double-counted version because GW231123, if unlensed, is an outlier in the tail of the astrophysical distribution with the highest masses and spins throughout the population.
If we used the leave-one-out version, the distribution would be extrapolating its tail at the parameters of unlensed GW231123, which would affect the posterior odds sensitively according to the functional form of the model at the tail.
By choosing to use the double-counted version of the astrophysical prior, we choose to err on the conservative side.

In our PE runs, we find that the lensed models give lower masses, lower spins and higher luminosity distances, all favored by the astrophysical prior.
Consequently, as shown in Fig.~\ref{fig:astro_prior}, reweighting from the uninformative prior to the astrophysical prior would mostly benefit the lensed hypothesis.
The one parameter whose reweighting would hurt the lensed hypothesis is the mass ratio $q$, as its inferred value is lower for the lensed hypothesis while the astrophysical prior has more weight for mass ratios closer to one.

\subsection{Lensed parameter astrophysical prior}\label{subsec:lens_astro}

The GWTC-5 population distribution described in Sec.~\ref{subsec:source_astro} only includes the source BBH parameters, and we still need to specify an astrophysical prior for the lens parameters $M_{Lz}$ and $y$ for all lens models, as well as $k$ for the gSIS model and $x_c$ for the CIS model.

As discussed at the beginning of Sec.~\ref{sec:astro_prior}, we should treat the normalization of the lensed parameter priors carefully.
If we use a normalization such that the prior probability integrates to one within our prior bounds, we would have effectively assumed that lensed events are as likely as unlensed ones a priori, which contradicts the fact that none of the other detected GW events were found to be lensed confidently.

For the gSIS model, the astrophysical prior of the lensed parameters can be written as
\begin{equation} \label{eq:lensed_prior}
    \tilde{\pi}(\vec{\theta}_{\rm L} | z_S) = \tilde{\pi}(M_{Lz}, y, k | z_S).
\end{equation}
For brevity, we will consider only the gSIS model in our discussion here, but the CIS case can be recovered simply by replacing $k$ with $x_c$, and the PM case by removing $k$ as appropriate and setting $\pi(k) = 1$ in subsequent steps.
To quantify the astrophysical prior of the lensed parameters, we ask the following question:
for a given source BBH at redshift $z_S$, what is the prior probability that there exists a lens with lens mass between $M_{Lz}$ and $M_{Lz} + dM_{Lz}$ lying between an impact parameter of $y$ and $y + dy$?
We can expand the three-dimensional prior,
\begin{multline}
        \tilde{\pi}(M_{Lz}, y, k|z_S) \\= \pi(y | y < y_{\rm cr})\tilde{\pi}(M_{Lz}, y < y_{\rm cr} | k, z_S) \pi(k), \label{eq:pi_k}
\end{multline}
where we choose $y_{\rm cr} \equiv 2$ in this work.
Here, $\pi(y | y < y_{\rm cr})$ integrates to one in the interval $0 < y <y_{\rm cr}$ by design, but the full prior $\tilde{\pi}(M_{Lz}, y, k | z_S)$ integrates to the prior probability of lensing for a source at $z_S$, which is lower than one in general (again, the tilde on top of $\pi$ is a reminder that it is not normalized within the prior bounds).
For simplicity, we assume that the prior in the lens parameter $k$ (or $x_c$) does not depend on $z_S$.
The shape of $\pi(y | y < y_{\rm cr})$ can be fixed by geometrical considerations, and does not depend on $z_S$.
On the source plane, the two-dimensional area $dA$ covered by a change $dy$ follows $dA \propto y \,dy$, so
\begin{equation}
    \pi(y | y < y_{\rm cr}) \propto y.
\end{equation}
The next term $\tilde{\pi}(M_{Lz}, y < y_{\rm cr} | k, z_S)$ is the prior probability (given a specific value of $k$) that a lens with redshifted lens mass $M_{Lz}$ exists within an impact parameter $y_{\rm cr}$ from the source.
The distribution of lenses on the celestial sphere is a spatial Poisson process, so 
\begin{equation}
    \tilde{\pi}(M_{Lz}, y< y_{\rm cr}|k, z_S) \approx \dfrac{d\tau_{y_{\rm cr}}}{dM_{Lz}}(k, z_S),
\end{equation}
where $\tau_{y_{\rm cr}}$ is the optical depth and the approximation holds when $\tau_{y_{\rm cr}} \ll 1$.
Explicitly,
\begin{equation} 
     \dfrac{d\tau_{y_{\rm cr}}}{dM_{Lz}}(k, z_S)  = 
     \int_0^{z_S} d z_L  \, \pi \left(\dfrac{\xi_0 y_{\rm cr}}{D_L}\right)^2 \dfrac{dn}{dM_{Lz}}\dfrac{1}{4\pi}\dfrac{dV_c}{dz_L},   \label{eq:opt_depth}
\end{equation}
where $n$ is the number density of lenses and $V_c$ is the comoving volume. 
The differential comoving volume $dV_c/dz_L$ is fixed by specifying the cosmology, which we set to be the Planck 2018 $\Lambda$CDM cosmology~\cite{Planck:2018vyg}.
The only quantities that depend on the lens model and its parameters ($k$ or $x_c$ in our case) are $\xi_0$ and $dn/dM_{Lz}$.
The magnitude of $\tilde{\pi}$ thus depends on $dn/dM_{Lz}$, the abundance of lenses in each $M_{Lz}$ bin.
This agrees with our intuition: if we believe that there are more lenses a priori, the prior probability of the lensing hypothesis is higher.
In the remainder of this section, we quantify $dn/dM_{Lz}$ for different scenarios and lens models.

\subsubsection{Point-mass lens}
\label{subsec:PM_astro}

\begin{figure}
    \centering
    \includegraphics[width=0.99\linewidth]{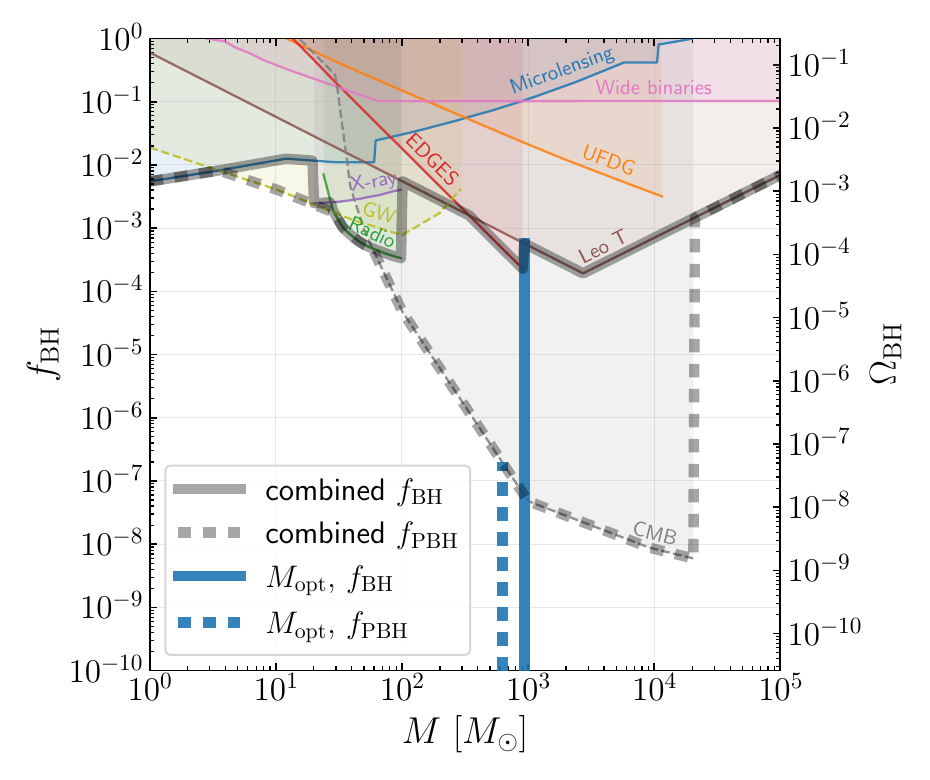}
    \caption{Observational constraints on the fraction of dark matter attributable to general BHs or PBHs, $f_{\rm BH}(M)$, as a function of the black hole mass $M$, assuming a monochromatic mass function.
    The thin curves show the upper limit from individual channels, with the region above each curve excluded.
    The PBH-only channels (CMB accretion and the GW merger rate) are drawn dashed, while the remaining channels apply to BHs of any origin.
    The two thick gray curves are the combined envelopes (the channel-wise minimum) that we feed into the optical depth calculation: the solid curve ($f_{\rm BH}$) drops the PBH-only bounds and represents the full BH population, while the dashed curve ($f_{\rm PBH}$) includes them and represents the PBH-only case.
    The vertical blue lines mark the optimal lens mass $M_{\rm opt}$ that maximizes the posterior odds under each envelope for the \texttt{NRSur7dq4} posterior (see Sec.~\ref{subsec:PM_astro}).
    The right axis shows the corresponding dark matter density fraction $\Omega_{\rm BH} = f_{\rm BH}\,\Omega_{\rm DM}$.
    The constraint compilation is taken from Ref.~\cite{Carr:2021bzv} and references therein.
    }
    \label{fig:f_BH}
\end{figure}

\begin{figure*}
    \centering
    \includegraphics[width=0.49\linewidth]{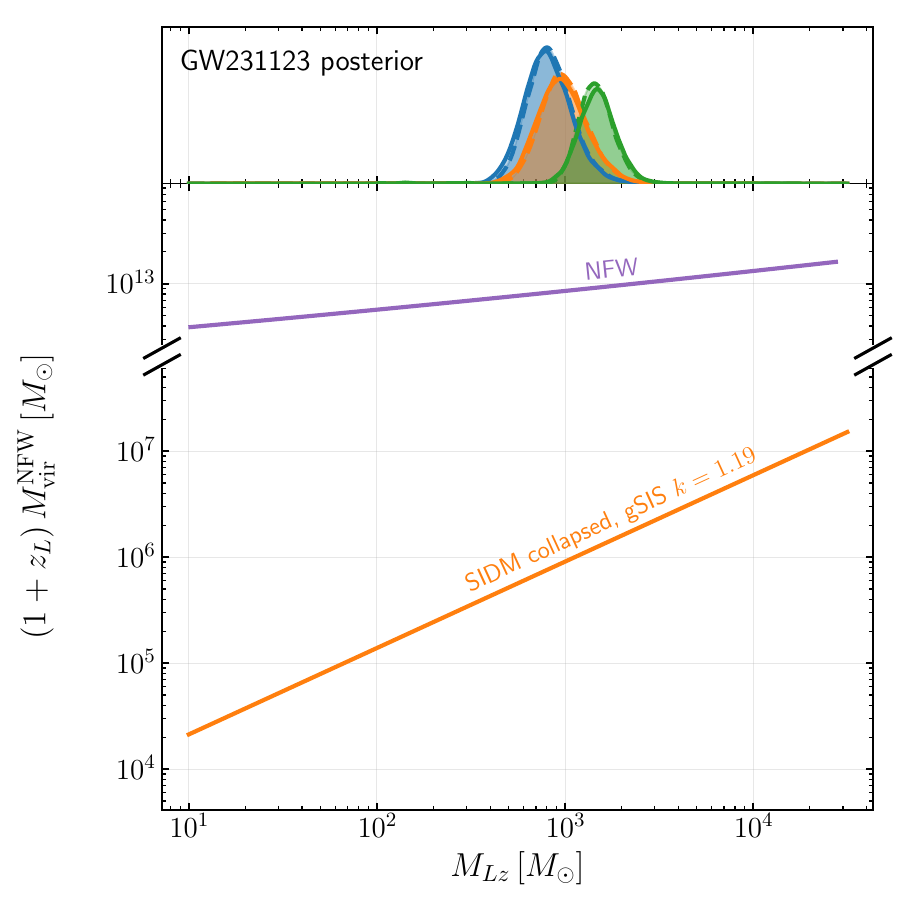}
    \includegraphics[width=0.49\linewidth]{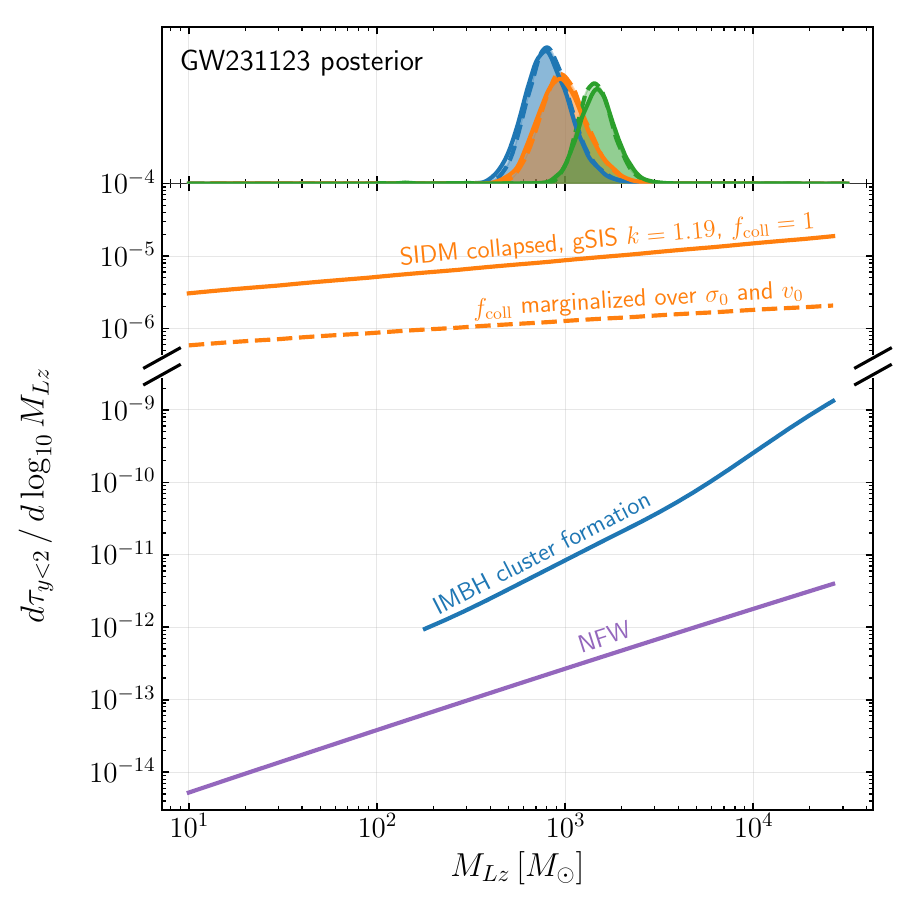}
    \caption{
    Relationship between the virial mass $M^{\rm NFW}_{\rm vir}$, redshifted lens mass $M_{Lz}$, and the differential optical depth $d\tau_{y<2}/d\log_{10}M_{Lz}$, for halos in collisionless CDM (NFW, purple curves) and those gravothermally collapsed in SIDM (orange curves).
    We also show the differential optical depth for IMBH lenses formed in star clusters (right panel, blue curve).
    \textit{Left:} $(1 + z_L)M^{\rm NFW}_{\rm vir}$ as a function of $M_{Lz}$, fixing $z_L = 0.4$ and $z_S = 0.8$.
    For a regular NFW halo in collisionless CDM, $M_{Lz}$ is orders of magnitude smaller than $M^{\rm NFW}_{\rm vir}$ because the inner profile $\rho(r) \sim r^{-1}$ is gentle.
    For a halo in SIDM that started as an NFW halo with $M^{\rm NFW}_{\rm vir}$ and later forms a collapsed core, $M_{Lz}$ is much closer to $M^{\rm NFW}_{\rm vir}$ because the inner profile $\rho(r)\sim r^{-2.19}$ is steep, allowing more effective mass to contribute to the lensing effects.
    These collapsed halos can be modeled by the gSIS profile with $k = 1.19$ for lensing purposes.
    \textit{Right:} The differential optical depth $d\tau_{y<2}/d\log_{10}M_{Lz}$ as a function of $M_{Lz}$ for a source at redshift $z_S = 0.8$, with the lens redshift integrated over $0 < z_L < z_S$.
    For IMBH lenses formed in star clusters, the differential optical depth is low, $\lesssim 10^{-9}$. 
    For the collisionless CDM case, as $M^{\rm NFW}_{\rm vir}$ is higher, the abundance of these halos is lower according to the halo mass function, which along with the small Jacobian $dM^{\rm NFW}_{\rm vir}/dM_{Lz}$ gives rise to an even lower optical depth.
    For the SIDM case, we show the differential optical depth if we assume that all halos would collapse ($f_{\rm coll} = 1$, solid orange line) and if we compute $f_{\rm coll}$ by marginalizing over the parameters governing the functional form of $\sigma(v)$ (dashed orange line), both orders of magnitude higher than the IMBH and NFW cases.
    }
    \label{fig:opt_depth}
\end{figure*}

For the point-mass lens model, it is difficult to quantify $dn/dM_{Lz}$ because there are no well-established mass functions for intermediate-mass black holes (IMBHs) that are calibrated to observations.
Nonetheless, there are observational constraints on the abundance of black holes throughout the mass spectrum, so we can impose an upper bound on the posterior odds.
These constraints come from microlensing observations, dynamical heating of ultra-faint dwarf galaxies (UFDGs), imprints of primordial black hole (PBH) accretion on the cosmic microwave background (CMB), GW merger rates due to PBHs, $21$-$\rm cm$ absorption heating constraints by the EDGES experiment, gas-cloud heating constraints from the dwarf galaxy Leo T, disruption of wide binary star systems in the Galactic halo, and X-ray and radio constraints of accretion near the Galactic center, see Ref.~\cite{Carr:2021bzv} for a full discussion.
In particular, both the CMB and GW constraints are on the abundance of PBHs in the early universe.
As any black hole could act as a lens no matter how it is formed, using the PBH bounds might over-constrain the abundance of PM lenses, as they could form through other astrophysical channels at a later epoch.
Therefore, we compute upper bounds on the posterior odds separately for the cases with and without the PBH-only constraints, i.e.\ those obtained from CMB and GW observations.
The bound without PBH-only constraints is a more lenient upper bound that allows for the possibility of a significant number of BHs forming astrophysically, while the one including PBH-only constraints assumes that all BHs are PBHs.

In the literature, constraints on the fraction of dark matter attributable to black holes $f_{\rm BH}(M)$ are usually obtained assuming that all BHs have the same mass, i.e.\ they follow a monochromatic distribution $\delta(M^\prime -M)$ and they have the same spatial distribution as dark matter.
In Fig.~\ref{fig:f_BH}, we show the observational constraints on $f_{\rm BH}(M)$ and $f_{\rm PBH}(M)$ for various channels.
At our mass range of interest $M \sim 10^3 \, {\rm M_\odot}$, whether or not to keep the CMB constraints would significantly affect our results.

To use the monochromatic $f_{\rm BH}$ constraints directly, we assume that all BHs have mass $M$, i.e.,
\begin{equation}
    \dfrac{dn}{dM'} = f_{\rm BH}(M) \dfrac{\rho_{{\rm DM},0}}{M}\delta(M' - M),
\end{equation}
where $\rho_{{\rm DM},0}$ is the present-day comoving dark matter density.
The differential optical depth in Eq.~\eqref{eq:opt_depth} becomes
\begin{equation} \label{eq:BH_opt_depth}
\dfrac{d\tau_{y_{\rm cr}}}{dM_{Lz}}  = 
     \pi\left.\left(\dfrac{\xi_0 y_{\rm cr}}{D_L}\right)^2 f_{\rm BH}(M) \dfrac{\rho_{{\rm DM},0}}{M^2}\dfrac{1}{4\pi}\dfrac{dV_c}{dz_L}\right|_{z_L^\prime},
\end{equation}
for $0 < z^\prime_L < z_S$, where $z_L^\prime \equiv M_{Lz}/M - 1$.
At values of $M_{Lz}$ where $z^\prime_L < 0$ or $z^\prime_L > z_S$, $d\tau_{y_{\rm cr}}/dM_{Lz} = 0$ because there is no way to get the required $M_{Lz}$ from $M$.
This scheme applies the same way to PBH bounds by replacing $f_{\rm BH}$ with $f_{\rm PBH}$.

To obtain a meaningful upper bound on $\mathcal{O}^{\rm PM}_{\rm unL}$ that respects the constraints, we maximize it over $M$ numerically, i.e.\ we find the maximum allowed value of $\mathcal{O}^{\rm PM}_{\rm unL}$ that respects the monochromatic $f_{\rm BH}(M)$ constraint at an optimal mass $M_{\rm opt}$.
For the \texttt{NRSur7dq4} posterior, the optima for the $f_{\rm PBH}$ and $f_{\rm BH}$ constraints are $M_{\rm opt} \approx \MoptfPBH \, {\rm M_\odot}$ and $\MoptfBH \, {\rm M_\odot}$ respectively.
Reweighting the source parameters to the GWTC-5 prior and using the differential optical depth in Eq.~\eqref{eq:BH_opt_depth} in the unnormalized lens parameter prior, we find the posterior odds for the PM model to be $\log_{10} \mathcal{O}^{\rm PM}_{\rm unL} = \PMfPBHopt$ and $\PMfBHopt$ for $f_{\rm PBH}$ and $f_{\rm BH}$ respectively.
Again, these numbers should be taken as optimal upper bounds that exhaust the constraints.

While there have been multiple proposed IMBH candidates, there has been no confident unambiguous detection of an IMBH with $M_\mathrm{BH}\gtrsim 10^3 M_\odot$ to this date \cite{Greene:2019vlv}, which is why we used observational constraints as an upper bound to the IMBH mass function above.
Other than using observational constraints, we can use a predicted IMBH mass function computed through simulations.
We use the mass function presented in Fig.~7 of Ref.~\cite{Kritos:2024sgd}, which is computed by semi-analytic simulations of stellar collisions and repeated BH mergers in nuclear star clusters (NSCs). 
Another possibility could be to use the BH mass function by convolving a known galaxy property distribution (such as the galaxy velocity dispersion function or bulge luminosity function) with extrapolation from an empirical scaling relation (such as $M_\mathrm{BH}$--$\sigma_*$ or  $M_\mathrm{BH}$--$M_\mathrm{bulge}$) \cite{Greene:2019vlv}, but we do not find a significant change in the corresponding optical depth.
IMBHs could also form in globular clusters (GCs) but with lower rates, so the mass function due to NSCs is a good order-of-magnitude approximation \cite{Kritos:2026qgk, Greene:2019vlv}.
Plugging the mass function as $dn/dM$ into Eq.~\eqref{eq:opt_depth} gives the differential optical depth, which we plot in the right panel of Fig.~\ref{fig:opt_depth}.
As the predicted abundance of IMBHs is low, the optical depth is low, giving low posterior odds of $\log_{10} \mathcal{O}^{\rm PM}_{\rm unL} = \PMIMBHcluster$. 
We believe this number reflects our prior belief more than those obtained with the $f_{\rm BH}$ and $f_{\rm PBH}$ constraints because it comes from astrophysical predictions that respect our current understanding about the formation mechanisms of IMBHs.
The numbers obtained from the constraints should only be treated as an optimistic upper bound that maximally respects current non-observation.

\subsubsection{Collisionless CDM halos}
\label{subsec:NFW_astro}

Numerical simulations have shown that collisionless cold dark matter (CDM) halos follow the NFW density profile across a broad range of halo masses.
However, as we show below, an NFW halo with concentrations fixed at realistic values is in the single-image weak-lensing regime and is not expected to be a good explanation of the data.
This is the reason why we do not perform our analysis with an NFW model.

The NFW lens has a density profile given by~\cite{Navarro:1995iw}
\begin{equation}
    \rho(r) = \dfrac{\rho_s}{r/r_s ( 1 + r/r_s)^2},
\end{equation}
where $r_s$ has units of length and $\rho_s$ has units of density.
We define the characteristic convergence~\cite{Bartelmann:1996hq}
\begin{equation}\label{eq:kappa_s}
    \kappa_s = \dfrac{\rho_s r_s}{\Sigma_{\rm crit}},
\end{equation}
where $\Sigma_{\rm crit} = c^2 D_S / 4 \pi G D_L D_{LS} $.
It is customary to define a concentration parameter $c$ such that 
\begin{equation} \label{eq:NFW_c_rs}
    R_{\rm vir} = c\, r_s ,
\end{equation}
where $R_{\rm vir} \equiv R_{200{\rm m}}$ is the radius within which the average density of the halos is $200$ times the mean matter density $\rho_{\rm m}$.
Then, it can be shown that 
\begin{equation}
    \rho_s = \delta_c \rho_{\rm m},
\end{equation}
where 
\begin{equation} \label{eq:NFW_delta_c}
    \delta_c = \frac{200}{3}  \dfrac{c^3}{\ln (1 + c) - c/(1+c)}.
\end{equation}
Fitting formulae for $c$ calibrated to cosmological $N$-body simulations exist in the literature, and we can use them to quantify the expected $\kappa_s$ given $M_{\rm vir}$ of the halo.
We use the fits in Ref.~\cite{Ishiyama:2020vao} implemented in the \texttt{Colossus} package~\cite{Diemer:2017bwl}.

In the literature, $\xi_0$ is usually chosen to be $r_s$ to simplify calculations.
However, such an $r_s$ could differ from $\xi_E$ by orders of magnitude, causing $y \sim O(1)$ to no longer be the threshold between negligible and significant lensing effects.
In general, the Einstein radius $\xi_E$ has to be solved numerically for the NFW lens.
For $\kappa_s \ll 1$, it can be approximated by
\begin{equation}
    \xi_E \approx 2 e^{-1/2} \, r_s \exp\left(-\dfrac{1}{2\kappa_s} \right),
\end{equation}
which approaches zero rapidly as $\kappa_s \to 0$.
As $\kappa_s \propto( M^{\rm NFW}_{\rm vir})^{0.21}$, $\xi_E$ drops from $\sim 1 \, \rm pc$ at $M^{\rm NFW}_{\rm vir} \sim 10^{13} \, \rm M_\odot$ to practically zero for $M^{\rm NFW}_{\rm vir} \lesssim 10^{12} \, {\rm M_\odot}$, see Fig.~\ref{fig:nfw_fits}.
As a result, a lens mass in the vicinity of $M_{Lz} \sim 10^3 \, \rm M_\odot$ requires a fine-tuned value of $M^{\rm NFW}_{\rm vir} \sim 6\times10^{12} \, \rm M_\odot$ on the galactic scale, see the left panel of Fig.~\ref{fig:opt_depth}.
The requirement of a fine-tuned $M^{\rm NFW}_{\rm vir}$ and the lower abundance of high mass halos imply that $dn/dM_{Lz}$ will be low for the NFW halo in our mass range of interest.
Indeed, as shown in the right panel of Fig.~\ref{fig:opt_depth}, the differential optical depth is $\gtrsim 7$ orders of magnitude lower than that for the gSIS as a collapsed SIDM halo, which we discuss below in Sec.~\ref{subsec:SIDM_astro}.
While we did not perform PE for the NFW model because of its low optical depth, we can estimate the order of magnitude of the posterior odds for the NFW model by assuming that the evidence integral with the uninformative priors is similar to that of the gSIS model.
Then, we can rescale the gSIS posterior odds to get $\log_{10}\mathcal{O}^{\rm NFW}_{\rm unL} \lesssim -6$.

\subsubsection{Prompt cusp}

\begin{figure*}
    \centering
    \includegraphics[width=0.525\linewidth]{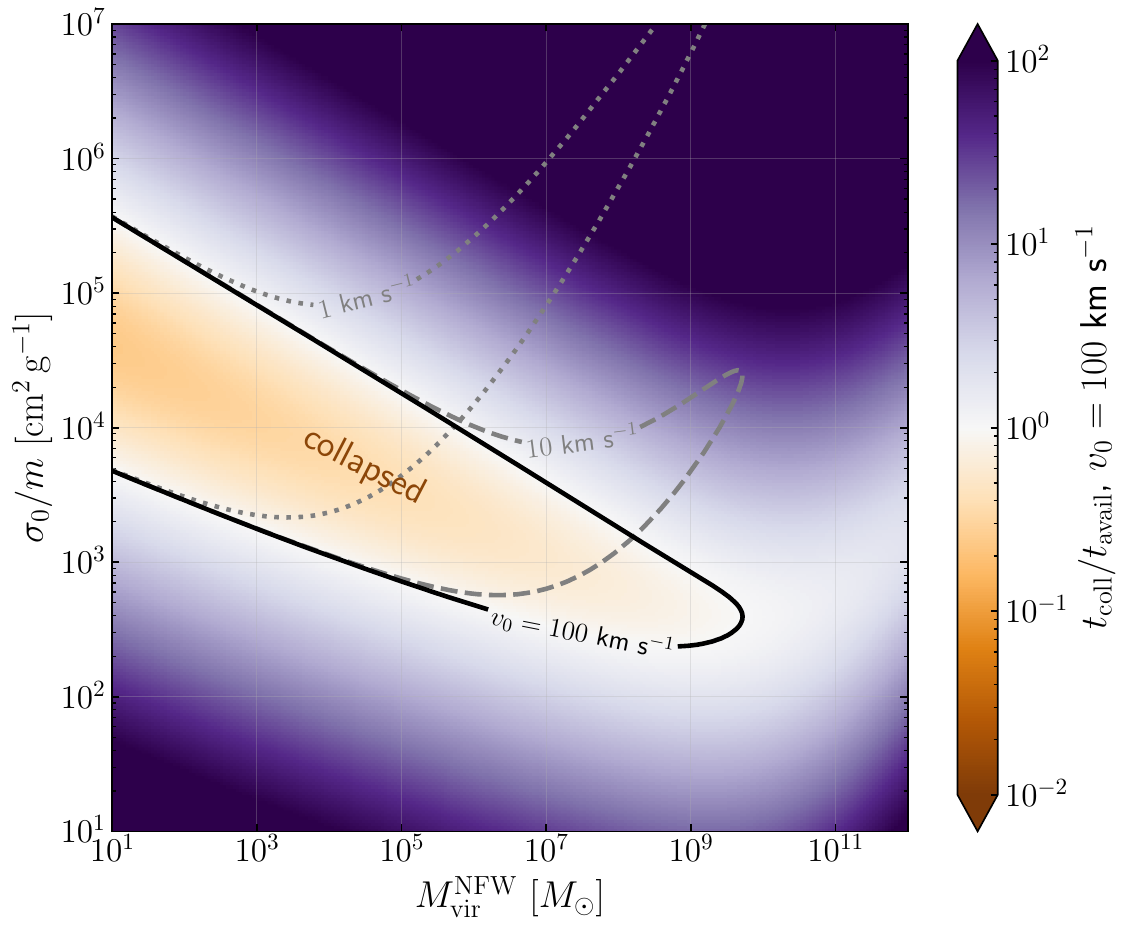}
    \includegraphics[width=0.465\linewidth]{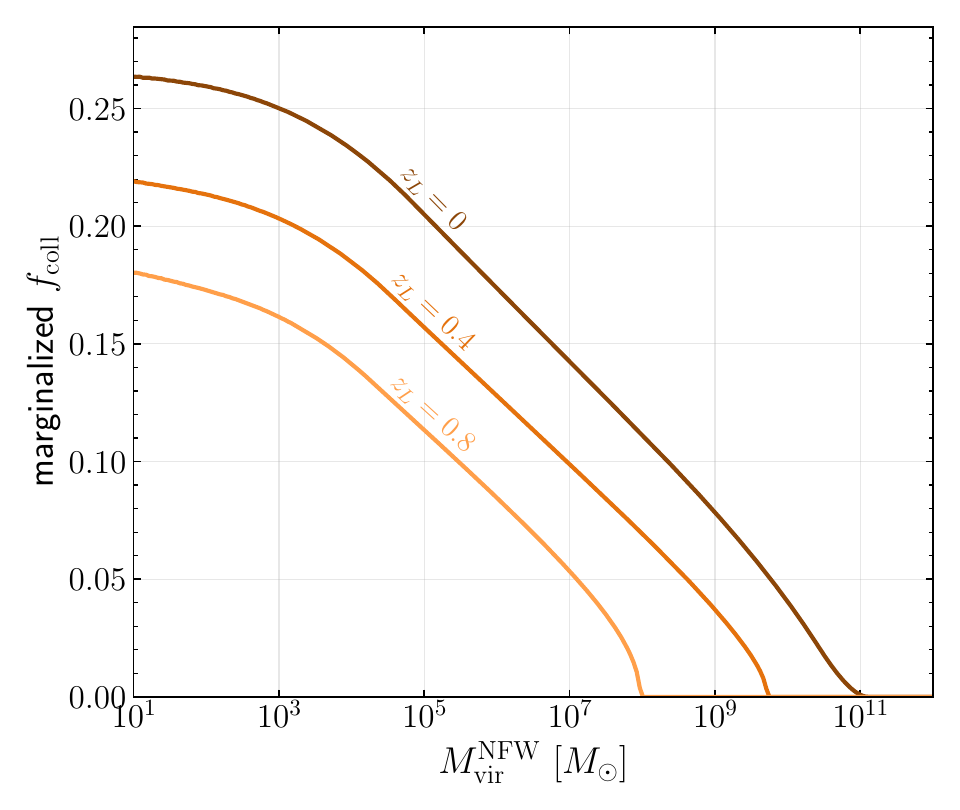}
    \caption{
    Whether an SIDM halo at redshift $z_L$ has collapsed gravothermally depends on the interaction cross section $\sigma(v)$, which is controlled by the parameters $\sigma_0$ and $v_0$ in our model (Eq.~\eqref{eq:sigma_v}).
    \textit{Left:} the time $t_{\rm coll}$ required for the halo to collapse, over the time $t_{\rm avail}$ available between the formation redshift $z_f$ and $z_L = 0.4$, as a function of $M^{\rm NFW}_{\rm vir}$ and $\sigma_0/m$, assuming $v_0 = 100 \, \rm km \, s^{-1}$.
    We bound the region $t_{\rm coll} < t_{\rm avail}$ with a black contour line, within which SIDM halos would have collapsed at $z_L = 0.4$. 
    We plot the same collapse region for $v_0 = 10 \, \rm km \, s^{-1}$ (gray dashed line) and $1 \, \rm km \, s^{-1}$ (gray dotted line).
    \textit{Right:} The fraction $f_{\rm coll}$ of SIDM halos that would have collapsed by redshift $z_L$, shown for $z_L = 0$, $0.4$ and $0.8$, if we marginalize over $\sigma_0/m \in [10^{-1}, 10^6] \, \rm cm^2 \, g^{-1}$ and $v_0 \in [10^{-1}, 10^2] \, \rm km \, s^{-1}$, both log-uniform.
    }
    \label{fig:sidm}
\end{figure*}

The density profile of DM halos at the very center can be cuspier than the NFW profile due to the formation of a prompt cusp at the moment of first collapse~\cite{Delos:2022yhn}, which could boost lensing effects.
However, for CDM, the mass of the cusp is $\sim 20$ orders of magnitude lower than the virial mass of the NFW halo~\cite{Delos:2025pen}, so even with a prompt cusp, CDM halos are not expected to be able to cause the strong lensing effects we might have observed.
For warm dark matter (WDM)~\cite{Bode:2000gq} with particle mass as high as $O(10) \rm keV$~\cite{Delos:2023exh}, the prompt cusp of $M_{\rm vir} \sim 10^{10} M_\odot$ halos could in principle have a lens mass compatible with our posterior in Fig.~\ref{fig:lens_params_posterior}.
However, the profile of the cusp was found to be $\rho(r) \sim r^{-1.5}$, i.e.\ $k \sim 0.5$ in our parameterization of the gSIS lens.
As the posterior of $k$ has no support there, we conclude that the prompt cusp lensing case is unlikely to explain the GW231123 observation.

\subsubsection{Gravothermally collapsed SIDM halos}
\label{subsec:SIDM_astro}

Initially, interest in self-interacting dark matter (SIDM)~\cite{Spergel:1999mh} was motivated by mismatches between collisionless CDM simulations and observational data on small scales, including the core-cusp problem and the missing satellite problem.
While the missing satellite problem has largely been resolved~\cite{Kim:2017iwr,DES:2019ltu,Sales:2022ich}, other tensions have arisen, including the too-big-to-fail problem and the diversity problem of galaxy rotation curves, which a velocity-dependent self-interaction can alleviate by thermalizing the inner halo into a core~\cite{Kaplinghat:2015aga,Tulin:2017ara,Adhikari:2022sbh}.
More recently, SIDM has also been invoked to explain observations that point to unexpectedly dense, centrally concentrated dark matter substructure, which is difficult to produce in collisionless CDM but arises naturally once a halo undergoes gravothermal core collapse.
These include the compact perturber inferred in the GD-1 stellar stream~\cite{Zhang:2024fib} and the dense substructures revealed by galaxy- and cluster-scale strong lensing~\cite{Minor:2020hic,Meneghetti:2020yif}, for which gravothermal collapse has been proposed as an explanation~\cite{Gilman:2021sdr,Zhang:2024fib}.
This is the same mechanism that we invoke below to form the compact halo lenses that could explain GW231123.

While a non-zero interaction cross section $\sigma$ can help resolve tensions with observations in lower mass systems, $\sigma$ has been well constrained in cluster and group scale systems with velocity dispersion $\gtrsim 100 \, \rm km \, s^{-1}$~\cite{Markevitch:2003at,Clowe:2003tk,Sagunski:2020spe,Andrade:2020lqq}.
This naturally motivates a velocity-dependent cross section $\sigma(v)$ that increases with decreasing velocity.
For example, a Yukawa interaction potential gives rise to a cross section
\begin{equation} \label{eq:sigma_v}
    \sigma(v) = \dfrac{\sigma_0}{(1 + (v/v_0)^2)^2}
\end{equation}
in the Born approximation~\cite{Ibe:2009mk,Tulin:2013teo}, so $\sigma(v) \to \sigma_0$ as $v\to0$, and $\sigma(v)\propto v^{-4}$ as $v  \to \infty$, with $v_0$ setting the velocity scale for transitioning between the two limits. 
In that case, lower mass halos with velocity dispersion $\lesssim v_0$ will have a non-negligible $\sigma$, and could undergo gravothermal collapse to form a cusp with $k \gtrsim 1$~\cite{Turner:2020vlf}.
We assume that the self-interacting nature of SIDM does not affect structure formation, so halos start off with NFW profiles.
Then, heat flows from the hot outer halo to the cold inner cusp, causing it to flatten into a constant-density core.
Later, the core reaches its maximum size and the direction of heat flow reverses, and the core loses energy and collapses into a steep cusp~\cite{Balberg:2002ue}.

While $\sigma(v)$ has to vary with velocity, the evolution of the core follows an approximately universal relation with an effective $\sigma_{\rm eff}$ that is velocity independent (but dependent on the properties of the halo through $v_{\rm max}$)~\cite{Outmezguine:2022bhq,Yang:2022zkd},
\begin{equation}
    \sigma_{\rm eff} = \dfrac{3}{2}\left.\dfrac{\langle\sigma_{\rm visc} v_{\rm rel}^3\rangle}{\langle v_{\rm rel}^3\rangle}\right|_{v = 0.64 v_{\rm max}},
\end{equation}
where $\sigma_{\rm visc} = \int d\sigma \sin^2\theta$ is the viscosity cross section, and $v_{\rm rel} = |\vec{v}_1 - \vec{v}_2|$ is the magnitude of the relative velocity between the interacting particles.
Angular brackets denote averaging over 3D Maxwell--Boltzmann distributions over both $\vec{v}_1$ and $\vec{v}_2$ with a 1D dispersion given by $v = 0.64 \,v_{\rm max}$, where $v_{\rm max} \approx 1.65 \,r_s \sqrt{G \rho_s}$ is the maximum rotational velocity within an NFW halo. 
The collapse time follows a universal relation~\cite{Yang:2022zkd,Mace:2025fuz}
\begin{equation}\label{eq:SIDM_universal}
    \beta_{\rm eff} \,\hat{\sigma}\,\hat{t}_{\rm coll} \approx 173,
\end{equation}
where $\hat{\sigma} = (\sigma_{\rm eff}/m) \rho_s r_s$ is the dimensionless cross section, with $m$ the mass of the SIDM particle, $\hat{t}_{\rm coll} = \sqrt{4 \pi G \rho_s}\, t_{\rm coll}$ is the dimensionless collapse time, and $\beta_{\rm eff}$ is the effective conduction coefficient, which follows a fitting function calibrated to $N$-body simulations~\cite{Mace:2025fuz},
\begin{equation}\label{eq:beta_eff}
    \beta_{\rm eff}(\hat{\sigma}) = \left( B^{- A} +\left(\dfrac{3.15}{\hat{\sigma}^2}\right)^{-A}\right)^{-1/A},
\end{equation}
where $B = 0.949$ and $A = 0.899$.
Then, given a $\hat{\sigma}$ and the formation redshift and $M_{\rm vir}$ of a halo, we can obtain the concentration $c$ from fits and get $r_s$ and $\rho_s$ with Eqs.~\eqref{eq:NFW_c_rs} to~\eqref{eq:NFW_delta_c}, and use Eqs.~\eqref{eq:SIDM_universal} and~\eqref{eq:beta_eff} to check whether the core has collapsed at a lower redshift, e.g. at $z_L$.
In Fig.~\ref{fig:sidm} we show $t_{\rm coll}$ as a function of $M_{\rm vir}$ and $\sigma_0/m$ for selected values of $v_0$.
For lower $M_{\rm vir}$ halos, they can collapse within a Hubble time if $\sigma(v)$ is within an allowed range at their $v_{\rm max}$. 

The collapsing core has been shown to evolve self-similarly through most of its lifetime~\cite{Balberg:2002ue,Outmezguine:2022bhq,Mace:2026gxg}, where the core radius $r_c$ shrinks and the profile stays as $\rho(r) \sim r^{-2.19}$ for $r_c \lesssim r \lesssim r_s$.
Explicitly, at late times~\cite{Mace:2026gxg},
\begin{equation}
    \rho(r) \approx \dfrac{\rho_c}{1 + (r/r_c)^\alpha( 1 + r/r_{\rm out})^{3-\alpha}},
\end{equation}
with $\alpha = 2.19$ and $r_{\rm out} \sim 2.9 \, r_s$.
As the core collapses, $r_c \to 0$ and $\rho_c \to \infty$, but the combination $\zeta_c = \rho_cr_c^\alpha$ is finite and $\sim 0.28 \rho_sr_s^\alpha$ close to collapse~\cite{Mace:2026gxg}.
For $r \ll r_{\rm out}$, 
\begin{equation} \label{eq:SIDM_halo_profile_approx}
    \rho(r) \approx \dfrac{\rho_c}{1 + (r / r_c)^\alpha},
\end{equation}
which is the profile for the CIS lens in Eq.~\eqref{eq:CIS_profile}, although with $\alpha = 2.19$ instead of $2$.
If we further require $r_c \ll r$, which is justified close to the end point of collapse $r_c \to 0$, 
\begin{equation} \label{eq:SIDM_halo_profile_approx_2}
    \rho(r) \approx \rho_c\left(\dfrac{r_c}{r}\right)^\alpha,
\end{equation}
which is the profile of the gSIS lens in Eq.~\eqref{eq:gSIS_profile} with $\rho_0 = \rho_c$, $r_0 = r_c$ and $k = \alpha - 1 = 1.19$.
In the lensing length scale $\xi_0$, $\rho_0$ and $r_0$ only appear as the combination $\rho_0r_0^{k + 1} = \zeta_c \sim 0.28 \rho_s r_s^\alpha$, giving $\xi_0/r_{\rm out} \lesssim 10^{-2}$ in our mass range, justifying the approximation $r \ll r_{\rm out}$, i.e.\ the lensing effects are dominated by the profile of the collapsing core instead of the outer NFW envelope.

While the inner profile of the halo follows a universal $\sim r^{-2.19}$ profile with a core during the onset of the collapse, the profile at and after the end point of collapse is not known.
Our gSIS and CIS lens models capture this uncertainty:
the gSIS model with $k \in [0.1, 1.9]$ accounts for the case where the collapse ends with a cusp as steep as $\sim r^{-2.9}$, while the CIS model accounts for the case where the collapse halts with a core of radius $r_c$ and an outer profile $\sim r^{-2}$.
In addition, the CIS model also captures the profile of the SIDM halo during the onset of collapse, where the core radius $r_c$ is not negligible compared to $\xi_0$.

To quantify the priors for the SIDM halo lensing hypothesis, we need to quantify $dn/dM_{Lz}$ in Eq.~\eqref{eq:opt_depth}.
The steps for obtaining priors for the gSIS model are similar to those for the CIS model, so the following discussion assumes the gSIS model, and we point out how the CIS model is different later.
For a fixed functional form of $\sigma(v)$, i.e.\ for fixed $\sigma_0$ and $v_0$ in Eq.~\eqref{eq:sigma_v}, we can write
\begin{equation} \label{eq:gamma_coll}
    \dfrac{d n}{d M_{Lz}} = \gamma_{\rm coll}(M^{\rm NFW}_{\rm vir}, z_f, z_L, \sigma_0/m, v_0) \dfrac{d n_{\rm full}}{d M^{\rm NFW}_{\rm vir}} \dfrac{dM^{\rm NFW}_{\rm vir}}{d M_{Lz}},
\end{equation}
where $n_{\rm full}$ is the number density of all halos, no matter if collapsed or not, and $\gamma_{\rm coll} = 1$ if a halo with virial mass $M^{\rm NFW}_{\rm vir}$ formed at redshift $z_f$ would have collapsed by redshift $z_L$, and $\gamma_{\rm coll} = 0 $ otherwise.
For given $\sigma_0/m$ and $v_0$, the $\gamma$ factor tells us to only include those halos that would have collapsed, i.e.\ those with $t_{\rm coll}$ less than the time evolved between $z_f$ and $z_L$, which can be computed from Eq.~\eqref{eq:SIDM_universal}.
To get $z_f$ as a function of $M^{\rm NFW}_{\rm vir}$, we use a fitting function from Ref.~\cite{Correa:2015kia}.
As we assumed that the self-interacting nature of SIDM does not affect structure formation, $dn_{\rm full}/dM^{\rm NFW}_{\rm vir}$ is just the usual halo mass function (HMF) obtained from collisionless CDM cosmological simulations.
To get $dM^{\rm NFW}_{\rm vir}/dM_{Lz}$, we first obtain the concentration parameter $c$ as a function of $M^{\rm NFW}_{\rm vir}$ from fits, and get $\rho_s$ and $r_s$ from Eqs.~\eqref{eq:NFW_c_rs} to~\eqref{eq:NFW_delta_c}.
Close to the end point of collapse, we have $\rho_0 r_0^{2.19} \sim 0.28 \rho_s r_s^{2.19}$.
Then, for the case of $k = 1.19$ and a redshift of $z_L$, using Eq.~\eqref{eq:xi0_gSIS} fixes $\xi_0$ and $M_{Lz}$ of the gSIS lens.
We thus obtain a relationship between $M^{\rm NFW}_{\rm vir}$ and $M_{Lz}$ and can compute its derivative.
This scheme only works for $k = 1.19$ because it is the true value seen in gravothermal collapse simulations, and we do not know how to convert from $\{\rho_s, r_s\}$ to $\{\rho_0,r_0\}$ if $k$ takes another value after the end of the collapse.
However, we can still use the same optical depth for other values of $k$ as an approximation, i.e.,
\begin{equation}
    \tilde{\pi}(M_{Lz}, y < y_{\rm cr} | k, z_S) \approx \tilde{\pi}(M_{Lz}, y < y_{\rm cr} | k = 1.19, z_S),
\end{equation}
cf. Eq.~\eqref{eq:pi_k}, and
\begin{equation} \label{eq:opt_depth_1.19_approx}
    \dfrac{d n}{d M_{Lz}} \approx \left. \dfrac{d n}{d M_{Lz}} \right|_{k = 1.19}.
\end{equation}

As $\sigma(v)$ is unconstrained for $v \lesssim 10 \, \rm km \,s^{-1}$, there is no reason to assume that $\sigma_0/m$ and $v_0$ would take any specific fine-tuned value.
Therefore, we marginalize over $\sigma_0/m$ and $v_0$ to get
\begin{align}\label{eq:f_coll}
    \dfrac{d n}{d M_{Lz}} &= f_{\rm coll}(M^{\rm NFW}_{\rm vir}, z_f, z_L) \dfrac{d n_{\rm full}}{d M^{\rm NFW}_{\rm vir}} \dfrac{dM^{\rm NFW}_{\rm vir}}{d M_{Lz}}, \\
    f_{\rm coll} &= \int d (\sigma_0/m)\, dv_0 \, \pi(\sigma_0/m,v_0) \, \gamma_{\rm coll}
\end{align}
where we integrate over $\sigma_0/m \in [10^{-1},10^{6}] \, \rm cm^2 \, g^{-1}$ and $v_0 \in [10^{-1}, 10^2] \, \rm km \, s^{-1}$.
The prior $\pi(\sigma_0/m, v_0)$ is log uniform in both parameters, but we set it to zero for combinations of $\sigma_0/m$ and $v_0$ that give $\sigma(100\,{\rm km \, s^{-1}})/m > 10 \, {\rm cm^2 \, g^{-1}}$ so as to satisfy observational constraints.
The factor $f_{\rm coll}$ is between $0$ and $1$, penalizing our hypothesis if it requires fine-tuned models of $\sigma(v)$ to give the collapsed halos required. 
In Fig.~\ref{fig:sidm}, we show $f_{\rm coll}$ as a function of $M^{\rm NFW}_{\rm vir}$.

Thus, we have all the components for computing the optical depth and the posterior odds for the gSIS lens assuming that it is a gravothermally collapsed SIDM halo.
For the case of an $f_{\rm coll}$ marginalized over $\sigma_0/m$ and $v_0$, we get $\log_{10}\mathcal{O}^{\rm gSIS}_{\rm unL} = \gSISfm$, which marginally favors the model over the unlensed one.
If we instead assume that all halos in our mass range of interest would gravothermally collapse, i.e.\ assume $f_{\rm coll} = 1$ (equivalent to assuming that $\sigma_0/m \sim 10^3$ with $v_0 \gtrsim 10 \, \rm km \, s^{-1}$; see the left panel of Fig.~\ref{fig:sidm}), we get $\log_{10}\mathcal{O}^{\rm gSIS}_{\rm unL} = \gSISfone$, substantially favoring lensing. 
However, the $f_{\rm coll} = 1$ assumption requires a semi-fine-tuned functional form of $\sigma(v)$, and there is no reason to believe so a priori, so the marginalized $f_{\rm coll}$ results more faithfully represent our uncertain prior belief in the $\sigma(v)$ model.
The bounds we pick to marginalize $\sigma_0/m$ and $v_0$ over affect the posterior odds, but that is a fundamental issue about parameters whose ranges we have no prior knowledge about.

While the above discussion is for the gSIS model, the same scheme applies equally well to the CIS model. 
As mentioned previously, using the CIS model captures our uncertainty about the core's density profile after the end point of the collapse.
In principle, it even captures the profile during the onset of the collapse when the core has a significant size, so the $t_{\rm coll}$ required to reach the CIS regime should be less than that for gSIS, increasing the fraction $f_{\rm coll}$ of halos that would give the needed profile.
However, only $x_c \lesssim 0.5$ gives rise to strong lensing effects, which maps to a nearly fully collapsed core $r_c / r_s \lesssim 10^{-2}$, so the $t_{\rm coll}$ required remains close to that in Eq.~\eqref{eq:SIDM_universal}.
Therefore, we can reuse the $f_{\rm coll}$ and optical depth computed for the gSIS lens.
Note that the outer slope of the CIS lens is equivalent to a gSIS with $k = 1$, which is different from the $k = 1.19$ of a collapsing core.
Nonetheless, the two values are near each other so we can use the optical depth of the $k = 1.19$ case directly, in the spirit of Eq.~\eqref{eq:opt_depth_1.19_approx}.
Including the optical depth information drops the posterior odds to $\log_{10}\mathcal{O}^{\rm CIS}_{\rm unL} = \CISfm$ for the marginalized $f_{\rm coll}$ case, which marginally disfavors lensing.
If we assume $f_{\rm coll} = 1$, we obtain $\log_{10}\mathcal{O}^{\rm CIS}_{\rm unL} = \CISfone$, which substantially favors lensing, but for the same reasons discussed for the gSIS case, the marginalized case is a more faithful representation of our prior belief.

\section{Frequentist lensing probability}
\label{sec:rate}

\subsection{Probability of detection from O1 to O4b}

The posterior odds of Sec.~\ref{sec:astro_prior} compare the lensed and unlensed hypotheses against each other, but say nothing about whether a GW231123-like lensing event is plausible given the rates and expected properties of lensed events.
We therefore supplement the odds with a frequentist consistency check: given the optical depth of a lens population, how probable is it to detect and identify a GW231123-like lensing event?
This matters because GW231123 is massive even under the lensed interpretation ($m_1\sim100\,{\rm M}_\odot$, $m_2\sim50\,{\rm M}_\odot$), whereas the GWTC-5 source population model prefers lower masses.
If rare lensing events do occur, they are most likely drawn from the low-mass bulk unless selection effects preferentially promote high-mass systems.

For this purpose, we define 
\begin{equation}
  p^{\rm L}_{{\rm obs}\geq1} \;=\; 1 - \exp\!\left(-N^{\rm L}_{\rm obs}\,p_{\vec{\theta}}\right),
  \label{eq:p_obs}
\end{equation}
the probability that, from O1 to O4b, we detect at least one event that is astrophysically lensed, identifiable as lensed, and with properties as atypical as the lensed GW231123.
Detected lensing events are rare and approximately Poisson, so $P(\ge 1)=1-e^{-\mu}$ with mean $\mu=N^{\rm L}_{\rm obs}\,p_{\vec{\theta}}$.
The factorization separates $N^{\rm L}_{\rm obs}$, the expected number of detected and identified lensing events of any parameters, from $p_{\vec{\theta}}\in[0,1]$, the fraction of the detected lensed population with parameters lying further away from the bulk than lensed GW231123 (i.e.\ more extreme than it).
In the following subsections we show how we estimate this quantity.
We assume that only the LIGO Hanford and Livingston detectors are on during the observational period in question.

\subsection{The expected number of detected and identified lensing events}
\label{subsec:ndet}

\begin{figure*}
    \centering
    \includegraphics[width=0.99\linewidth]{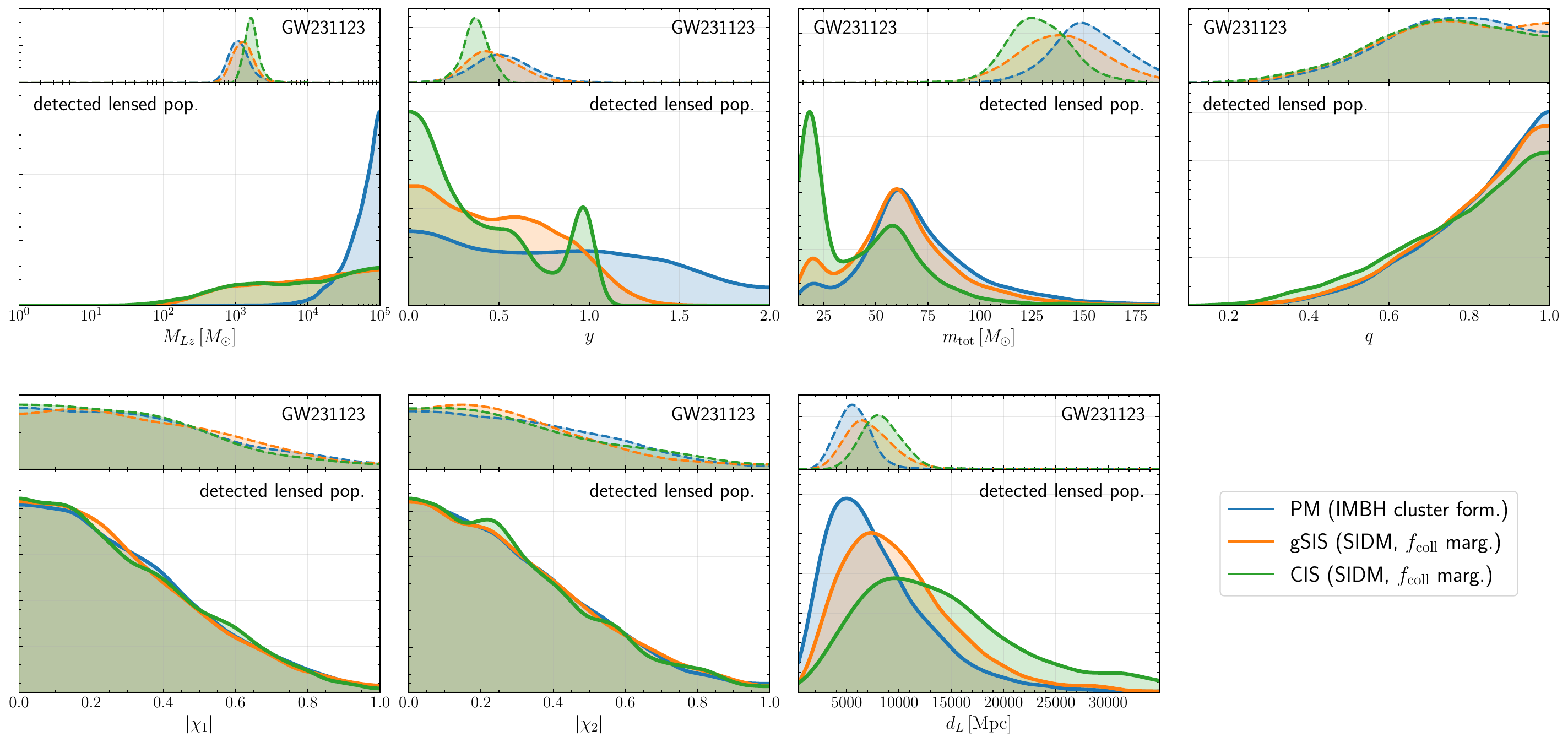}
    \caption{The astrophysically reweighted posterior of lensed GW231123 compared to the detected population of lensed events for the PM (IMBH star cluster formation prior), gSIS and CIS (both SIDM collapsed halo prior marginalized over $f_{\rm coll}$) models.
    The source parameter prior is always chosen to be the GWTC-5 population distribution.
    \textit{Top panels:} The posterior of lensed GW231123 reweighted to the corresponding astrophysical priors. 
    \textit{Bottom panels:} The normalized detected lensed population simulated assuming that the intrinsic population follows the astrophysical priors we imposed.
    Lensed GW231123 is in the tail of the population distribution of $m_{\rm tot}$ for all lens models, and in the tail of $M_{Lz}$ for the PM model with the IMBH cluster formation channel prior. 
    }
    \label{fig:det_pop}
\end{figure*}

In this section, we set $\pi(\vec{\theta}_{\rm S})$ to be the normalized, dimensionless GWTC-5 population distribution.
However, the population model uses $\propto (1+z)^\kappa$ for the redshift distribution, which is not calibrated well above $z \sim 1$, so we replace it with the shape of the Madau-Dickinson star-formation rate~\cite{Madau:2014bja}, which agrees with the population distribution for $z \lesssim 1$.

We assume that most if not all of these events are unlensed, i.e.\ $N^{\rm unL}_{\rm obs} \approx N_{\rm obs}$.
Then, for a given normalized intrinsic distribution of sources $\pi(\vec{\theta}_{\rm S})$,
\begin{equation}
    N^{\rm unL}_{\rm obs} = R_0 T \int d\vec{\theta}_{\rm S}\, \pi(\vec{\theta}_{\rm S})\,P^{\rm unL}_{\rm det}(\vec{\theta}_{\rm S}),
\end{equation}
where $P^{\rm unL}_{\rm det}(\vec{\theta}_{\rm S})$ is the probability that our search pipeline would be able to detect an event with false alarm rate ${\rm FAR} < 1 \, {\rm yr}^{-1}$ given its parameters $\vec{\theta}_{\rm S}$, and $R_0$ is the overall merger rate and $T$ the observing time, so that $R_0 T$ is the expected total number of mergers.
For lensed observations, we have
\begin{equation}
    N^{\rm L}_{\rm obs} = R_0 T \int d\vec{\theta}_{\rm L} d\vec{\theta}_{\rm S}\, \pi(\vec{\theta}_{\rm S})\,\tilde{\pi}(\vec{\theta}_{\rm L}|z_S)P^{\rm L}_{\rm det}(\vec{\theta})P^L_{\rm id}( \vec{\theta}),
\end{equation}
where $\vec{\theta} = \{\vec{\theta}_{\rm L},\vec{\theta}_{\rm S}\}$, and $\tilde{\pi}(\vec{\theta}_{\rm L} | z_S) = \tau(z_S) \pi(\vec{\theta}_{\rm L} | z_S)$ is the unnormalized distribution of lensed parameters that includes the optical depth in its magnitude, cf. Eq.~\eqref{eq:pi_lens}, while $P^{\rm L}_{\rm det}(\vec{\theta})$ is similar to $P^{\rm unL}_{\rm det}(\vec{\theta}_{\rm S})$ but for the lensed waveform. 
The additional term $P^L_{\rm id}( \vec{\theta})$ is the probability that the lensed waveform would be identified as lensed during parameter estimation.

For the observing runs $\{\text{O1}, \text{O2}, \text{O3a}, \text{O3b}, \text{O4a}, \text{O4b}\}$, we have detected $\{N_{\rm obs}\}=\{3, 8, 40, 25, 86, 104\}$ events with ${\rm FAR} < 1 \, {\rm yr}^{-1}$~\cite{LIGOScientific:2018mvr,LIGOScientific:2021usb,KAGRA:2021vkt,LIGOScientific:2021psn,LIGOScientific:2025slb,LIGOScientific:2026wfs}.
Then, for each observing run, and for each source and lens parameter prior setup (i.e.\ each row of Table~\ref{tab:odds_NRSur}), we can estimate the number of lensed detections by a Monte Carlo sum over a simulated intrinsic population of 
\begin{equation}
    N^{\rm L}_{\rm obs} = N^{\rm unL}_{\rm obs} \frac{\sum_{i, \pi({\vec{\theta}})} \tau(z_{S, i})\,P^{\rm L}_{{\rm det}}(\vec{\theta}_i)P^{\rm L}_{{\rm id}}(\vec{\theta}_i)}{\sum_{i, \pi(\vec{\theta}_{\rm S})} P^{\rm unL}_{{\rm det}}(\vec{\theta}_{{\rm S},i})}. \label{eq:N_obs_MC}
\end{equation}
The numerator is a Monte Carlo sum over events drawn from the normalized joint distribution $\pi(\vec{\theta}) \equiv \pi(\vec{\theta}_{\rm L}|z_S)\pi(\vec{\theta}_{\rm S})$, while the denominator is a sum over events drawn from $\pi(\vec{\theta}_{\rm S})$, with each draw labeled by $i$.

In the denominator, we use a $P^{\rm unL}_{{\rm det}}$ calibrated to published injection recovery campaign data of search pipelines~\cite{LIGOScientific:2021psn,Essick:2025zed}, which gives a more faithful representation of selection effects than a naive constant signal-to-noise ratio (SNR) detection threshold.
We approximate $P^{\rm unL}_{{\rm det}}(\vec{\theta}_{\rm S}) \equiv P^{\rm unL}_{{\rm det}} (\rho_{\rm mf}, m_{\rm det})$ as a function of $\rho_{\rm mf}$, the matched-filter SNR of the event, and $m_{{\rm det}}$, its detector-frame total mass.
Among the intrinsic parameters, only $m_{{\rm det}}$ is retained, in addition to the SNR, because $P^{\rm unL}_{{\rm det}}$ depends dominantly on it.
For unlensed events, we assume that the template bank represents all signals faithfully, so $\rho_{\rm mf} \approx \rho_{\rm opt}$, where $\rho_{\rm opt}$ is the optimal SNR. 

For the Monte Carlo integral in the numerator, the probability $P^{\rm L}_{{\rm det}}$ is the same as $P^{\rm unL}_{{\rm det}}(\rho_{{\rm mf},i}, m_{{\rm det},i})$ described above, but computed using the $\rho_{{\rm mf}, i}$ between the fully lensed waveform and the best-fit unlensed template, and using the $m_{{\rm det},i}$ of that template.
This reflects the mismatch between the signal and template when using unlensed template banks to search for lensed signals~\cite{Chan:2024qmb}.
In principle, as the lensed and unlensed waveforms could differ by a lot, search pipelines might veto a candidate event if it were lensed. 
This should be included in $P^{\rm L}_{{\rm det}}$ as a penalty for lens waveforms that deviate significantly from any unlensed one, e.g. by downweighting $\rho_{\rm mf}$ according to a $\chi^2$ signal consistency test~\cite{Allen:2004gu} like in the \texttt{pycbc} search pipeline~\cite{Usman:2015kfa}.
We include such a penalty in $P^{\rm L}_{{\rm det}}$, but find that it does not affect the results in this section qualitatively.

For $P^{\rm L}_{{\rm id}}$, we set it to take binary values $P^{\rm L}_{{\rm id}} = \{0, 1\}$, depending on whether the lensed signal is distinguishable from the best-fit unlensed waveform.
Following the indistinguishability criterion of Ref.~\cite{Toubiana:2024car}, and defining the fitting factor (FF) as the overlap between the lensed waveform and an optimal unlensed waveform maximized over time and phase, we deem event $i$ identifiable as lensed when the residual ${\rm SNR}^2$ left after subtracting the best-fit unlensed template exceeds the $90\%$ quantile of the $\chi^2(n_L)$ distribution,
\begin{equation}
    (1 - {\rm FF}_i^2)\,(\rho^L_{{\rm opt},i})^2 \;>\; \mathcal{Q}_{0.9}(n_L),
\end{equation}
where ${\rm FF}_i$ is the fitting factor of event $i$, $\rho^L_{{\rm opt},i}$ is the optimal SNR of the lensed waveform, $n_L$ is the number of lens parameters for the model in question, and $\mathcal{Q}_{0.9}(n_L)$ is the $90\%$ quantile of a chi-squared distribution with $n_L$ degrees of freedom ($\mathcal{Q}_{0.9} = 4.61$ and $6.25$ for $n_L = 2$ and $3$).
Equivalently, this requires the true lensed parameters to lie outside the $90\%$ confidence region of the unlensed fit.

\subsection{Is GW231123 typical of the predicted lensed population?}
\label{subsec:pbulk}

The Monte Carlo integral above tells us how many lensed events we expect to detect given our lens population $\tilde{\pi}(\vec{\theta}_{\rm L}|z_S)$.
A combined value of $N^{\rm L}_{\rm obs} \sim 1$ over all observing runs implies that it is not surprising to see a lensed event in the data.
However, even if that is the case, it does not mean that we should expect to see a lensed event with the properties of GW231123. 
As mentioned previously, even if GW231123 were lensed, the source-frame masses are still high compared to the bulk of the GWTC-5 source population distribution.
If we only observed one lensed event, it would probably have been one with lower masses (see Ref.~\cite{Farah:2025ews} for a similar argument but for lensed events in the geometrical optics regime).

To quantify how often we would detect a lensed event with GW231123-like properties, we simulate a detected catalog of lensed events given our source and lens population distributions (Fig.~\ref{fig:det_pop}).
Such a population can be generated by sampling from $\pi(\vec{\theta})$ weighted by the integrand in the numerator of Eq.~\eqref{eq:N_obs_MC}.
Then, we work in the 7D parameter space $\{\log_{10} M_{Lz},\, y,\, m_{\rm tot},\, q,\, |\chi_{1}|,\, |\chi_{2}|,\, d_L\}$ and use a Gaussian kernel-density estimator (KDE) to estimate its 7D probability density function (PDF). 
We note that a KDE depends on the metric in the parameter space, and we choose to work in coordinates in which the 7D distribution has unit variance, i.e.\ whitened.
For each posterior sample of GW231123, we obtain the PDF at the corresponding parameter space location and compare it with all the events in the simulated population.
The fraction of events $p$ with a PDF less than that of the GW231123 sample tells us how rare the sample is.
If $p_{\vec{\theta}} \ll 1$, the sample lies in the tail of the detected distribution, meaning that it is inherently rare to have detected a lensed event with such parameters.
If $p_{\vec{\theta}} \sim 1$, the sample lies in the bulk of the detected distribution and is typical.

\subsection{Combined frequentist statistic}

As $N^{\rm L}_{\rm obs}$ is the expected number of lensed events detected assuming our source and lens population model, $N^{\rm L}_{\rm obs} \,p_{\vec{\theta}}$ is then the expected number of lensed events with properties as atypical as GW231123.
Using $N^{\rm L}_{\rm obs} \,p_{\vec{\theta}}$ instead of $N^{\rm L}_{\rm obs}$ allows us to quantify the atypicalness of observing a lensed GW231123-like event.
Moreover, $N^{\rm L}_{\rm obs}$ depends on the bounds of our lens population.
We bounded our population within $M_{Lz} \in [1, 10^5] \, \rm M_\odot$, but broadening the bounds might increase $N^{\rm L}_{\rm obs}$, especially when the upper bounds are increased (although diffraction effects will be less important, and the time delay could be larger than the length of the in-band signal itself, giving separate images).
As the lensed posterior of GW231123 has $M_{Lz} \sim 10^3 \, \rm M_\odot$, including $p_{\vec{\theta}}$ in the statistic offsets this effect, because although increasing the upper bound changes $N^{\rm L}_{\rm obs}$, it does not change the number of detected events with $M_{Lz} \lesssim 10^3 \, \rm M_\odot$, which is $N^{\rm L}_{\rm obs} \,p_{\vec{\theta}}$.

After converting $N^{\rm L}_{\rm obs} \,p_{\vec{\theta}}$ to $p^{\rm L}_{\rm obs \geq 1}$ with Eq.~\eqref{eq:p_obs}, we find that for all prior setups, $p^{\rm L}_{\rm obs \geq 1} \lesssim 10^{-3}$, meaning that a lensed GW231123-like event is inherently unlikely, see the right-most column of Table~\ref{tab:odds_NRSur}.

A similar quantity, $p^{\rm unL}_{\rm obs \geq 1}$, can be computed for the unlensed case, which tells us how likely we would have detected an unlensed GW231123-like event from O1 to O4b if the population follows the GWTC-5 distribution. 
We expect $p^{\rm unL}_{\rm obs \geq 1}$ to be $O(0.1)$ because GW231123 is included when inferring the GWTC-5 population distribution.
Indeed, through a similar procedure described in the previous subsections, we find $p^{\rm unL}_{\rm obs \geq 1} = \Punlensed$.
The reason why $\mathcal{O}^{\rm L}_{\rm unL}$ can be orders of magnitude different from $p^{\rm L}_{\rm obs \geq 1}/p^{\rm unL}_{\rm obs \geq 1}$ is because the former takes into account the difference in likelihood and the Occam penalty while the latter does not.
Note that the Occam penalty favors the lensed hypothesis even when it has more parameters because the posteriors for the source parameters are wider when a lensed model is used.

\section{Discussion} \label{sec:discussion}

\subsection{Is GW231123 diffraction-lensed?}

In Sec.~\ref{sec:uninformative}, we show that the lensed models fit better than the unlensed ones and give meaningfully measured lensed parameters.
However, when astrophysical priors are included in Sec.~\ref{sec:astro_prior}, the posterior odds do not favor the lensed hypothesis for the priors we deem to be most faithfully representative of our belief, i.e.\ an IMBH mass function predicted from cluster formation channels for PM, and a mass function of collapsed halo cores assuming SIDM for gSIS and CIS.
The posterior odds $\log_{10} \mathcal{O}^{\rm L}_{\rm unL} = \gSISfm$ and $\CISfm$ for gSIS and CIS seem to suggest that while not significantly favored, the lensed hypothesis is not ruled out.
However, as we show in Sec.~\ref{sec:rate}, the probability that we would detect a lensed event like GW231123 from O1 to O4b is low, with $p^{\rm L}_{\rm obs \geq 1} \lesssim 10^{-3}$.
Therefore, we conclude that GW231123 is unlikely to be diffraction-lensed by an isolated object.

The reason why $\mathcal{O}^{\rm L}_{\rm unL}$ does not rule out lensing for some model and prior choices while the frequentist probability $p^{\rm L}_{\rm obs \geq 1}$ significantly disfavors lensing is that $\mathcal{O}^{\rm L}_{\rm unL}$ compares the lensed hypothesis to the baseline unlensed one.
The unlensed hypothesis could very well be untrue or incomplete.
For example, the data could be contaminated by non-Gaussian noise transient glitches~\cite{Ray:2025rtt}, the binary might be eccentric~\cite{Jan:2025zcm,Chandra:2026voe}, the waveform models could have significant systematic errors in the parameter space of GW231123~\cite{LIGOScientific:2025rsn}, or there could be beyond-GR signatures in the signal~\cite{Liu:2026wor,Lai:2026yvm}.
A complete analysis should compare the posterior odds across all hypotheses with reasonable priors.
It would not be surprising to find that one of these hypotheses is significantly favored over both the lensed and unlensed hypotheses we explored here.

\subsection{Lens prior assumptions}
\label{sec:lens_discussion}

A major limitation of our work is that we assumed that the lens exists in isolation.
If the lens is embedded within the gravitational potential of a larger galaxy or cluster, the lensing effects will be enhanced and the morphology of the waveform will change nontrivially.
The optical depth calculation will also have to include the abundance of such embedded lenses.
Refs.~\cite{Goyal:2025eqo,Shan:2025dcd} found that embedded lensing models fit GW231123 well.
To the best of our knowledge, a posterior odds with astrophysical priors on the embedded lensing case for GW231123 has not been presented in the literature, and we defer such calculation to future work.
In fact, for the SIDM-motivated gSIS and CIS priors specifically, if the lens is a subhalo embedded in the tidal field of a host halo rather than existing in isolation, a larger fraction of such halos may undergo gravothermal core collapse, and collapse may be reached at a lower self-interaction cross section, since tidal stripping of the outer, higher-velocity-dispersion material steepens the halo's density profile and hastens core collapse (see e.g. Refs.~\cite{Nishikawa:2019lsc,Sameie:2019zfo,Kahlhoefer:2019oyt,Zeng:2023fnj}).

Other than assuming that they are isolated, we assumed that the lens can be well modeled by the spherically symmetric lens models we chose.
For the PM model, if the IMBH exists inside an NSC or a GC, the density profile should be dressed by a distribution of matter outside of the BH. 
For the gSIS model, we allow the slope parameter $k$ to vary because we want to capture the uncertainty of the end point of SIDM gravothermal collapse, but there is no reason why the profile would remain a clean power law.
The uniform prior imposed on $k$ might also penalize the hypothesis because it allows for values of $k$ away from the predicted $k = 1.19$ of gravothermal collapse, although the penalty would not change our conclusions qualitatively.
Similarly, for the CIS model, we effectively assumed that the lens profile can be well described by the CIS profile, and the uniform prior in $x_c$ might penalize the hypothesis, though not significantly.

Moreover, when using the SIDM gravothermally collapsed halo, we effectively assumed that SIDM is the correct model for DM. 
The posterior odds $\mathcal{O}^{\rm L}_{\rm unL}$ for the SIDM rows in Table~\ref{tab:odds_NRSur} should be multiplied by an additional factor in the prior odds that reflects our prior belief in whether DM is SIDM.
We do not include such a factor because to the best of our knowledge, there is no way to quantify our prior belief on whether DM is self interacting or not.
Therefore, if one believes that SIDM is extraordinary, then one should require $\mathcal{O}^{\rm L}_{\rm unL}$ to pass a high threshold to conclude that the lensed hypothesis is supported.
On top of this, we have also imposed a prior that is uniform on a logarithmic scale over $\sigma_0/m \in [10^{-1}, 10^6] \, \rm cm^2 \, g^{-1}$ and $v_0 \in [10^{-1}, 10^2] \, \rm km \, s^{-1}$, and setting it to zero for combinations of $\sigma_0/m$ and $v_0$ that give $\sigma(100\,{\rm km \, s^{-1}})/m > 10 \, {\rm cm^2 \, g^{-1}}$.
To the best of our knowledge, these prior ranges do not violate observational constraints, but changing their bounds or shapes will affect our results.

\subsection{Source prior assumptions}

There are also biases and uncertainties in the source parameter astrophysical prior that we did not account for in our analysis.
Strictly speaking, the GWTC-5 population distribution includes GW231123 itself when it was inferred, meaning that using such a distribution for analyzing GW231123 will lead to double counting.
Usually, the best practice is to perform population inference while leaving out the event in question, and then using such a leave-one-out population distribution as the prior for analyzing that event (see e.g., \cite{Moore:2021xhn}). 
However, we choose not to do so because GW231123 is an outlier with high masses and high spins lying in the tail of the population distribution.
If the event is left out from the population analysis, the astrophysical probability will effectively be extrapolating at its parameter values, so it will depend sensitively on how the tail of the distribution is parameterized, e.g. whether the tail is a power law or an exponential cutoff.
Doing this risks suppressing the astrophysical prior probability of seeing such an outlier event more than it should.
By using the full GWTC-5 population distribution inferred from events including GW231123, we choose to err on the conservative side (i.e.\ favoring the unlensed hypothesis).

\subsection{Comparison to previous literature}\label{sec:comparison}
 
Several recent works have analyzed GW231123 under the lensing or related
hypotheses~\cite{Goyal:2025eqo,LIGOScientific:2025cwb,Chan:2025kyu,Hu:2025lhv,Shan:2025dcd,Chakraborty:2025pxt,Wang:2026yjk,Barsode:2026bcs},
and our analysis is complementary to these.
The LVK GWTC-4.0 lensing search~\cite{LIGOScientific:2025cwb} analyzed GW231123 under two models: an isolated point-mass (PM) model and a fold caustic, with the simulated background drawn from the GWTC-4.0 source population.
Ref.~\cite{Goyal:2025eqo} extends the PM analysis to include an external anisotropic gravitational potential representing a strongly lensing host galaxy (the ``embedded PM'' model), reporting a false alarm probability ${<}\,1\%$.
Ref.~\cite{Chan:2025kyu} uses the \texttt{DINGO-lensing} framework to perform ${\sim}\,2\times10^{5}$ simulations under a PM lens hypothesis, quoting a frequentist false alarm probability that corresponds to a significance from $3$ to $4\sigma$ depending on the waveform model used.
Ref.~\cite{Hu:2025lhv} does not model GW231123 as a lensed event, but shows that a model with two overlapping BBH signals is favored over the
isolated-signal hypothesis, providing an independent alternative interpretation of the inter-waveform-model discrepancies in the standard unlensed analysis.
Ref.~\cite{Shan:2025dcd} considered a binary system embedded in a galaxy as the lens model, and found that the lensing hypothesis is favored with $\log_{10} \mathcal{B} = 2.67$.
Ref.~\cite{Chakraborty:2025pxt} tested for unmodeled correlated features between the residual strain of the two LIGO detectors, which could be a signature of microlensing effects, but found that waveform systematics might be sufficiently large to induce such features in the residual for this event.
Ref.~\cite{Wang:2026yjk} showed that diffraction-lensing effects are degenerate with signatures of a graviton with nonzero mass in this event.
Ref.~\cite{Barsode:2026bcs} uses the Bayes factor-Bayes factor relationship and quantifies the significance of the PM lensing hypothesis at $\lesssim 4.1 \sigma$.
 
Our analysis differs from these works in two respects.
First, in addition to the PM model, we consider extended halo lens
profiles, the gSIS model with a variable slope and the CIS model with a variable core radius (Sec.~\ref{subsec:lens_models}), which to the best of our knowledge have not previously been applied to GW231123.
The motivation is that if the lens is a dark matter halo rather than a compact object, its diffuse density profile changes both the diffraction signature and the prior probability of the configuration through the optical depth, which depends on the lens model used.
Second, the posterior odds we quote are simultaneously reweighted by astrophysical priors on the source parameters and on the lens parameters; see Sec.~\ref{sec:astro_prior}. This allows our quoted posterior odds to directly reflect the prior rarity of both the unlensed and lensed hypotheses.

\section{Conclusions}

In this work, we reanalyzed GW231123 under the hypothesis that it is diffraction-lensed by an isolated, spherically symmetric lens modeled as a PM, gSIS or CIS lens.
With uninformative priors, the lensed models fit the data better than the unlensed one, returning well-constrained lens parameters and source parameters that shift to lower, more common masses and spins.
This comparison is incomplete on its own, however, because the lensed hypothesis has more parameters and is a priori rare, so we quantified the astrophysical prior on the source parameters using the GWTC-5 population distribution and on the lens parameters using the lensing optical depth.
For the priors we deem most faithful to our astrophysical prior belief, the posterior odds do not favor lensing significantly.
For a PM lens, observational bounds on the BH abundance give only an upper bound on the odds that does not by itself rule out lensing ($\log_{10} \mathcal{O}^{\rm PM}_{\rm unL} = \PMfBHopt$), but a predicted mass function of IMBHs formed in clusters yields a low optical depth that strongly disfavors it ($\log_{10} \mathcal{O}^{\rm PM}_{\rm unL} = \PMIMBHcluster$).
For a halo lens, collisionless CDM strongly disfavors lensing ($\log_{10} \mathcal{O}^{\rm NFW}_{\rm unL} \lesssim -6$), while SIDM halos that have undergone gravothermal core collapse give inconclusive posterior odds ($\log_{10} \mathcal{O}^{\rm gSIS}_{\rm unL} = \gSISfm$ and $\log_{10} \mathcal{O}^{\rm CIS}_{\rm unL} = \CISfm$).
In all cases, a frequentist estimate shows that detecting a lensed event as atypical as GW231123 over O1 to O4b is unlikely, with $p^{\rm L}_{\rm obs \geq 1} \lesssim 10^{-3}$.
We therefore conclude that GW231123 is unlikely to be diffraction-lensed by an isolated object.
These conclusions rest on our assumptions about the shape and compactness of the lens as well as the DM model assumed for the halo case.
Moreover, they only apply to isolated lenses, since a lens embedded in an external gravitational potential could change both the morphology of the waveform and the optical depth.

When we have a better understanding of the abundance and density profile of low mass DM halos, our prior belief can be updated and our posterior odds could change.
With more observations of GW231123-like events we will also be able to more robustly infer the tail properties of the population distribution in the high mass and spin regime, allowing us to perform a leave-one-out estimation of the astrophysical prior.
We anticipate that a smoking gun for lensing would be a subpopulation of events with an inferred impact parameter $y$ following a distribution $P(y) \propto y$ (modulo selection effects), which is a clean, purely geometric prediction.

\section*{Data availability}

The data that support the findings of this article and the code for reproducing all results are publicly available~\cite{cheung_2026_21515053}.

\begin{acknowledgments}
We thank Srashti Goyal, Hector Villarrubia-Rojo and Miguel Zumalac\'{a}rregui for pointing out a mismatch in Fourier conventions in an earlier version of our code, as well as for many useful comments and discussion.
We thank Jose Mar\'{i}a Ezquiaga and Luka Vujeva for useful comments and discussion.
We thank the Institute for Advanced Study Astrophysics Group and the Johns Hopkins University Gravitational Wave Astrophysics Group for valuable comments and discussion.
M.H.Y.C. is a Croucher Fellow supported by the Croucher Foundation.
M.H.Y.C. and M.Z. acknowledge support from the Nelson Center for Collaborative Research at the Institute for Advanced Study and the Simons Foundation through Award No. SFI-MPS-BH-00012593-10.
M.Z. acknowledges support from the National Science Foundation NSF-BSF 2207583 and NSF 2209991. 
T.V. acknowledges support from NSF grant no. 2309360, the Alfred P. Sloan Foundation through grant number FG-2023-20470, and the BSF through award number 2022136. 
This work used Anvil~\cite{song2022anvil} at Purdue University through allocation PHY250264 from the Advanced Cyberinfrastructure Coordination Ecosystem: Services \& Support (ACCESS) program~\cite{boerner2023access}, which is supported by U.S. National Science Foundation grants No.~2138259, 2138286, 2138307, 2137603, and 2138296.
This work used large language models (LLMs) including \texttt{Claude Fable 5}, \texttt{Claude Opus 4.5}, \texttt{4.6}, \texttt{4.7} and \texttt{4.8}, \texttt{Claude Sonnet 4.6} and \texttt{5}, \texttt{Claude Haiku 4.5} and \texttt{Gemini 2.5 Pro} to assist with literature review, brainstorming, coding, executing data analysis scripts, and editing the manuscript, all with close supervision by the authors.
This work used the \texttt{astropy}~\cite{Astropy:2022ucr}, \texttt{bilby}~\cite{bilby_paper}, \texttt{colossus}~\cite{Diemer:2017bwl}, \texttt{corner}~\cite{corner}, \texttt{dynesty}~\cite{Speagle:2019ivv}, \texttt{gwpy}~\cite{Macleod:2021goi}, \texttt{hmf}~\cite{Murray:2013qza}, \texttt{jax}~\cite{jax2018github}, \texttt{lalsuite}~\cite{lalsuite}, \texttt{matplotlib}~\cite{Hunter:2007ouj}, \texttt{numpy}~\cite{Harris:2020xlr}, \texttt{pesummary}~\cite{Hoy:2020vys}, and \texttt{scipy}~\cite{Virtanen:2019joe} software packages.
\end{acknowledgments}

\appendix

\section{Uninformative prior and parameterization}
\label{app:priors}

For the source parameters defined in \texttt{bilby}, we use the following uninformative prior for all runs:

\begin{align}
  \pi(m_1) &\sim \text{Uniform}\label{eq:mc_gsIs}\\
  \pi(m_2)             &\sim \text{Uniform} \label{eq:q_gsis}\\
  \pi(|\chi_1|)           &\sim \mathcal{U}[0,\; 0.99] \label{eq:a1_gsis}\\
  \pi(|\chi_2|)           &\sim \mathcal{U}[0,\; 0.99] \label{eq:a2_gsis}\\
  \pi(\theta_1)      &\propto \sin\theta_1, \quad \theta_1 \in [0,\;\pi] \label{eq:tilt1_gsis}\\
  \pi(\theta_2)      &\propto \sin\theta_2, \quad \theta_2 \in [0,\;\pi] \label{eq:tilt2_gsis}\\
  \pi(\phi_{12})     &\sim \mathcal{U}[0,\; 2\pi] \label{eq:phi12_gsis}\\
  \pi(\phi_{JL})     &\sim \mathcal{U}[0,\; 2\pi] \label{eq:phijl_gsis}\\
  \pi(\theta_{JN})   &\propto \sin\theta_{JN}, \quad \theta_{JN} \in [0,\;\pi] \label{eq:thetajn_gsis}\\
  \pi(\psi)          &\sim \mathcal{U}[0,\;\pi] \label{eq:psi_gsis}\\
  \pi(\phi_c)        &\sim \mathcal{U}[0,\; 2\pi] \label{eq:phase_gsis}\\
  \pi(\kappa_z)      &\propto \sin\kappa_z, \quad \kappa_z \in [0,\;\pi]  \label{eq:zenith_gsis}\\
  \pi(\alpha_{\rm az}) &\sim \mathcal{U}[0,\; 2\pi] \label{eq:az_gsis}\\
  \pi(t_{H1})        &\sim \mathcal{U}[t_{\rm event}-0.1,\; t_{\rm event}+0.1]\;\text{s} \label{eq:time_gsis}.
\end{align}

While the prior is uniform in $m_1$ and $m_2$, for sampling we reparameterize to the chirp mass $\mathcal{M}$ and the mass ratio $q \equiv m_2/m_1 \leq 1$, with bounds $\mathcal{M} \in [40, 200] \, M_\odot$ and $q \in [1/6, 1]$ for the \texttt{NRSur7dq4} runs and $q \in [1/10, 1]$ for the \texttt{IMRPhenomXPHM-SpinTaylor} runs, the tighter bound reflecting the validated range of \texttt{NRSur7dq4}.
We additionally impose a constraint that the source-frame component masses lie within $m_1, m_2 \in [20, 250] \, M_\odot$.
For the luminosity distance $d_L$, we use a prior uniform in the comoving volume adjusted by a factor of $1/(1 + z)$ to account for the redshift of the source-frame merger rate (``uniform-source-frame'').
When sampling, we use this prior in $d_L$ directly for the unlensed runs, but use a slightly different prior for the lensed runs as explained below and reweight to the same uniform-source-frame prior immediately after sampling.

For the lens parameters,
\begin{align}
    \pi(M_{Lz}) &\sim \text{LogUniform}, \quad M_{Lz} \in [10, 5\times10^5] \, M_\odot \\
    \pi(y) &\propto y, \quad y \in [0.01, 2],
\end{align}
but for sampling we reparameterize $M_{Lz}$ and $d_L$ to $\Delta{\tilde{T}}$ and $\tilde{d}_L$,
\begin{align}
    \Delta\tilde{T} &= 8 M_{Lz} \, y \\
    \tilde{d}_L &= d_L \, y,
\end{align}
so that the two parameters are less degenerate with $y$.
For an SIS lens ($k = 1$ for gSIS, $x_c = 0$ for CIS), $\Delta{\tilde{T}}$ is exactly equal to the time delay between the two images in the geometrical optics limit.
Its bounds are taken to be $\Delta{\tilde{T}} \in [0.1, 2\times10^4] \, GM_\odot/c^3 $.
For the lensed runs, during sampling, the prior used for $\tilde{d}_L$ is a power law with index $\alpha = 1/2$ with bounds $\tilde{d}_L \in [1, 4000] \, {\rm Mpc}$, but reweighted to uniform-source-frame in $d_L$ after sampling.
All the results quoted for the uninformative prior for the lensed runs have been reweighted this way.
For the additional lens parameters, for gSIS,
\begin{equation}
    \pi(k) \sim \mathcal{U}[0.1, 1.9],
\end{equation}
and for CIS,
\begin{equation}
    \pi(x_c) \sim \mathcal{U}[0, 0.5].
\end{equation}

For all runs, similar to the original GW231123 analysis~\cite{LIGOScientific:2025rsn}, we also marginalize over the detector calibration uncertainties by sampling $10$ spline nodes per detector for the amplitude and phase, drawn from the LVK calibration envelopes.

We perform the sampling with different numbers of live points $N_{\rm live}$ to check that the sampler has converged to the correct posteriors and evidence integrals.
However, we find that the maximum-likelihood parameters are not compatible across nested sampling runs even with $N_{\rm live} \sim 10^4$, which is close to being prohibitively expensive computationally.
This pathology affects both the lensing and unlensed hypotheses.
The likelihood surface seems to contain narrow peaks that are difficult to sample efficiently.
This effect is being investigated in an ongoing work, and might be related to the edge-on and highly precessing nature of GW231123 (see Appendix~\ref{app:insights}).
Nonetheless, this does not affect our conclusions in this work as the posterior and evidence integrals have converged, and we are not drawing any conclusions from the maximum likelihood values, except in Fig.~\ref{fig:waveform} for visualization of the waveform.

\section{Full parameter estimation results}

In the main text, we analyzed GW231123 with two waveform models and four lens models.
In Table~\ref{tab:posterior_summary}, we show a summary of the posterior distributions of the model parameters for reference.
The full posterior samples are publicly available~\cite{cheung_2026_21515053}.

\begin{table*}[htbp]
\centering
\begin{tabular}{lcccc}
\toprule
  & \textbf{unL} & \textbf{PM} & \textbf{gSIS} & \textbf{CIS} \\
\midrule
  \multicolumn{5}{l}{\textbf{IMRPhenomXPHM-SpinTaylor}} \\
\midrule
  $m_1\ [M_\odot]$ & $149.0^{+14.2}_{-13.1}$ & $118.9^{+28.0}_{-26.4}$ & $111.2^{+28.7}_{-29.3}$ & $98.0^{+28.4}_{-24.6}$ \\[0.6em]
  $m_2\ [M_\odot]$ & $91.1^{+20.9}_{-21.9}$ & $52.6^{+26.1}_{-21.8}$ & $48.5^{+25.8}_{-20.0}$ & $43.9^{+23.7}_{-17.6}$ \\[0.6em]
  $\mathcal{M}\ [M_\odot]$ & $118.6^{+11.8}_{-14.5}$ & $113.1^{+18.0}_{-30.9}$ & $112.1^{+18.6}_{-30.3}$ & $111.6^{+18.2}_{-30.1}$ \\[0.6em]
  $q$ & $0.61^{+0.13}_{-0.14}$ & $0.45^{+0.30}_{-0.22}$ & $0.45^{+0.31}_{-0.21}$ & $0.46^{+0.32}_{-0.22}$ \\[0.6em]
  $\chi_\mathrm{eff}$ & $0.02^{+0.18}_{-0.25}$ & $0.24^{+0.26}_{-0.30}$ & $0.23^{+0.26}_{-0.30}$ & $0.20^{+0.28}_{-0.31}$ \\[0.6em]
  $\chi_p$ & $0.72^{+0.20}_{-0.24}$ & $0.57^{+0.33}_{-0.40}$ & $0.58^{+0.33}_{-0.42}$ & $0.55^{+0.35}_{-0.41}$ \\[0.6em]
  $|\chi_1|$ & $0.78^{+0.18}_{-0.27}$ & $0.70^{+0.24}_{-0.47}$ & $0.71^{+0.24}_{-0.51}$ & $0.69^{+0.26}_{-0.55}$ \\[0.6em]
  $|\chi_2|$ & $0.68^{+0.28}_{-0.58}$ & $0.50^{+0.44}_{-0.45}$ & $0.49^{+0.45}_{-0.44}$ & $0.49^{+0.45}_{-0.44}$ \\[0.6em]
  $\cos\theta_1$ & $-0.27^{+0.40}_{-0.37}$ & $0.46^{+0.49}_{-0.67}$ & $0.45^{+0.49}_{-0.68}$ & $0.43^{+0.51}_{-0.78}$ \\[0.6em]
  $\cos\theta_2$ & $0.64^{+0.33}_{-1.03}$ & $0.23^{+0.71}_{-1.06}$ & $0.18^{+0.75}_{-1.02}$ & $0.17^{+0.75}_{-1.02}$ \\[0.6em]
  $\theta_{JN}$ & $1.56^{+0.44}_{-0.42}$ & $0.84^{+1.96}_{-0.58}$ & $0.94^{+1.88}_{-0.67}$ & $0.87^{+1.96}_{-0.64}$ \\[0.6em]
  $d_L\ [\mathrm{Mpc}]$ & $880^{+422}_{-341}$ & $4004^{+2272}_{-1827}$ & $4800^{+3541}_{-2212}$ & $6459^{+3377}_{-2902}$ \\[0.6em]
  $M_{Lz}\ [M_\odot]$ & --- & $892.3^{+663.6}_{-308.4}$ & $1066.2^{+765.5}_{-381.3}$ & $1523.8^{+740.5}_{-422.3}$ \\[0.6em]
  $k$ & --- & --- & $1.42^{+0.43}_{-0.55}$ & --- \\[0.6em]
  $x_c$ & --- & --- & --- & $0.03^{+0.05}_{-0.03}$ \\[0.6em]
  $y$ & --- & $0.62^{+0.30}_{-0.26}$ & $0.53^{+0.27}_{-0.21}$ & $0.40^{+0.13}_{-0.12}$ \\[0.6em]
  $\ln\mathcal{L}$ & $209.7^{+4.1}_{-5.8}$ & $215.7^{+3.6}_{-5.2}$ & $215.6^{+3.7}_{-5.2}$ & $215.3^{+3.7}_{-5.2}$ \\
\midrule
  \multicolumn{5}{l}{\textbf{NRSur7dq4}} \\
\midrule
  $m_1\ [M_\odot]$ & $127.4^{+16.9}_{-17.2}$ & $116.4^{+33.1}_{-30.0}$ & $107.5^{+36.7}_{-30.8}$ & $92.9^{+31.7}_{-23.0}$ \\[0.6em]
  $m_2\ [M_\odot]$ & $105.5^{+16.3}_{-18.8}$ & $55.2^{+26.9}_{-23.4}$ & $50.5^{+27.0}_{-21.6}$ & $48.1^{+22.9}_{-19.7}$ \\[0.6em]
  $\mathcal{M}\ [M_\odot]$ & $138.3^{+5.8}_{-12.0}$ & $117.7^{+17.4}_{-29.7}$ & $116.4^{+18.1}_{-30.2}$ & $117.1^{+17.7}_{-27.9}$ \\[0.6em]
  $q$ & $0.83^{+0.14}_{-0.16}$ & $0.48^{+0.35}_{-0.25}$ & $0.48^{+0.35}_{-0.26}$ & $0.53^{+0.33}_{-0.27}$ \\[0.6em]
  $\chi_\mathrm{eff}$ & $0.32^{+0.18}_{-0.34}$ & $0.33^{+0.24}_{-0.33}$ & $0.32^{+0.25}_{-0.35}$ & $0.26^{+0.29}_{-0.35}$ \\[0.6em]
  $\chi_p$ & $0.74^{+0.17}_{-0.18}$ & $0.47^{+0.37}_{-0.33}$ & $0.47^{+0.39}_{-0.33}$ & $0.45^{+0.41}_{-0.33}$ \\[0.6em]
  $|\chi_1|$ & $0.91^{+0.07}_{-0.26}$ & $0.71^{+0.25}_{-0.55}$ & $0.70^{+0.26}_{-0.55}$ & $0.64^{+0.32}_{-0.54}$ \\[0.6em]
  $|\chi_2|$ & $0.91^{+0.08}_{-0.30}$ & $0.49^{+0.45}_{-0.44}$ & $0.50^{+0.44}_{-0.45}$ & $0.49^{+0.45}_{-0.44}$ \\[0.6em]
  $\cos\theta_1$ & $0.73^{+0.23}_{-0.82}$ & $0.68^{+0.29}_{-0.84}$ & $0.66^{+0.31}_{-0.86}$ & $0.61^{+0.35}_{-1.09}$ \\[0.6em]
  $\cos\theta_2$ & $-0.05^{+0.48}_{-0.41}$ & $0.21^{+0.71}_{-1.06}$ & $0.23^{+0.70}_{-1.06}$ & $0.20^{+0.72}_{-1.03}$ \\[0.6em]
  $\theta_{JN}$ & $1.23^{+0.90}_{-0.34}$ & $0.72^{+2.10}_{-0.46}$ & $0.91^{+1.92}_{-0.63}$ & $0.80^{+2.03}_{-0.54}$ \\[0.6em]
  $d_L\ [\mathrm{Mpc}]$ & $2015^{+1668}_{-1084}$ & $4376^{+2389}_{-1780}$ & $5327^{+3640}_{-2349}$ & $7075^{+3252}_{-2977}$ \\[0.6em]
  $M_{Lz}\ [M_\odot]$ & --- & $872.9^{+672.1}_{-317.6}$ & $1054.8^{+816.1}_{-399.6}$ & $1560.8^{+795.9}_{-477.0}$ \\[0.6em]
  $k$ & --- & --- & $1.43^{+0.42}_{-0.57}$ & --- \\[0.6em]
  $x_c$ & --- & --- & --- & $0.03^{+0.06}_{-0.03}$ \\[0.6em]
  $y$ & --- & $0.64^{+0.34}_{-0.27}$ & $0.54^{+0.30}_{-0.23}$ & $0.39^{+0.15}_{-0.12}$ \\[0.6em]
  $\ln\mathcal{L}$ & $212.4^{+4.4}_{-5.9}$ & $215.5^{+4.2}_{-5.3}$ & $215.6^{+4.1}_{-5.4}$ & $215.2^{+4.0}_{-5.3}$ \\
\bottomrule
\end{tabular}
\caption{Summary of the posterior distribution of parameters inferred for all waveform and lens model combinations used in the main text.
The numbers shown are the median and $90\%$ credible interval.
The component masses $m_1$ and $m_2$ are in the source frame, while the chirp mass $\mathcal{M}$ is in the detector frame.
The results shown in this table use the uninformative prior.}
\label{tab:posterior_summary}
\end{table*}

\section{Further insights on the analysis of GW231123}
\label{app:insights}

In this appendix, we discuss some further observations regarding how our models fit the GW231123 data. 

\begin{figure*}
    \centering
    \includegraphics[width=0.99\linewidth]{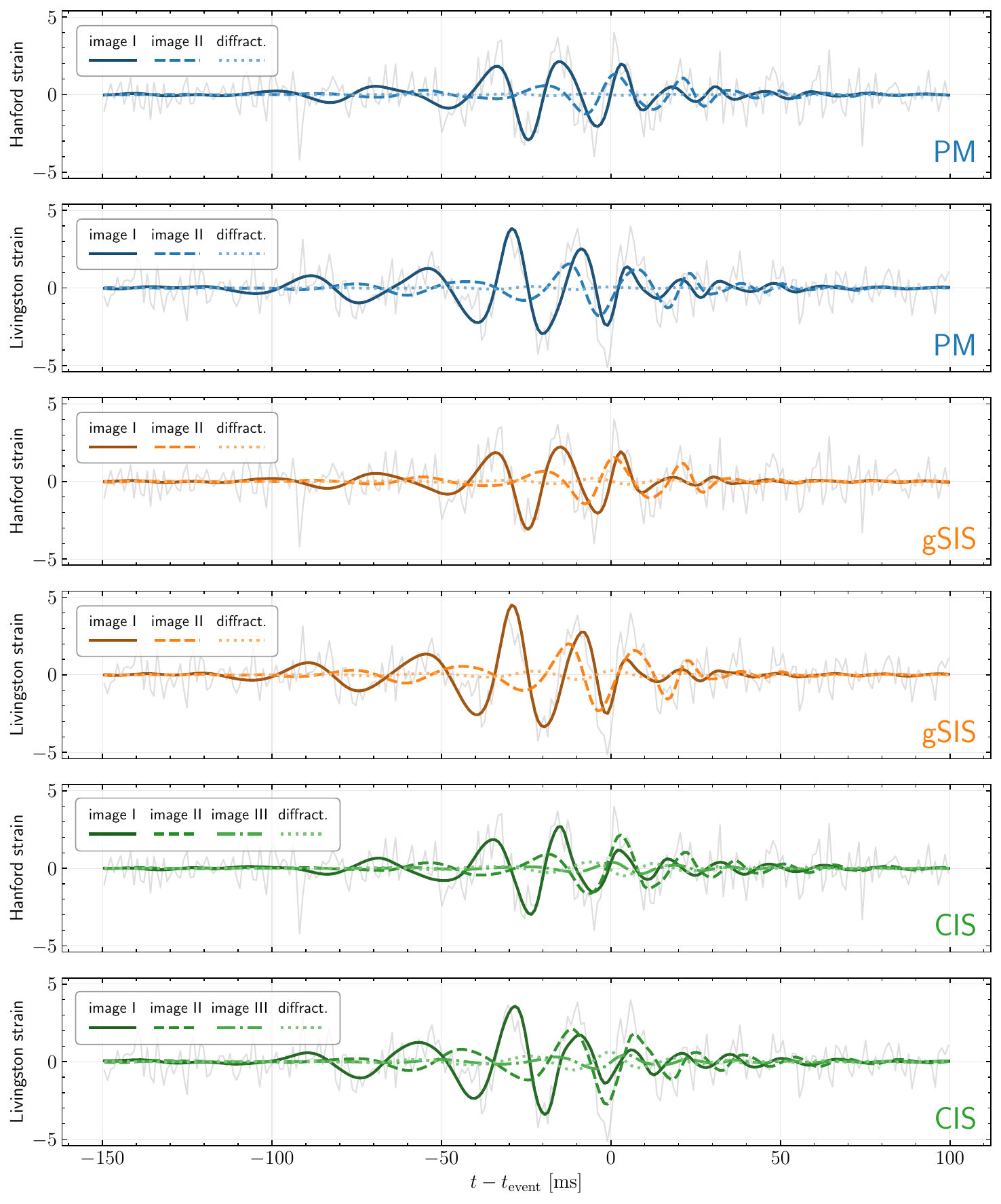}
    \caption{The maximum likelihood waveform for the PM (top two), gSIS (middle two) and CIS (bottom two) lens models for the Hanford and Livingston detectors, decomposed into geometrical optics limit images and a diffraction component.
    The waveform model used is \texttt{NRSur7dq4}.
    The magnification and time delay of each component are shown in Fig.~\ref{fig:mag_delta_T}.
    }
    \label{fig:images}
\end{figure*}

\begin{figure*}
    \centering
    \includegraphics[width=0.6\linewidth]{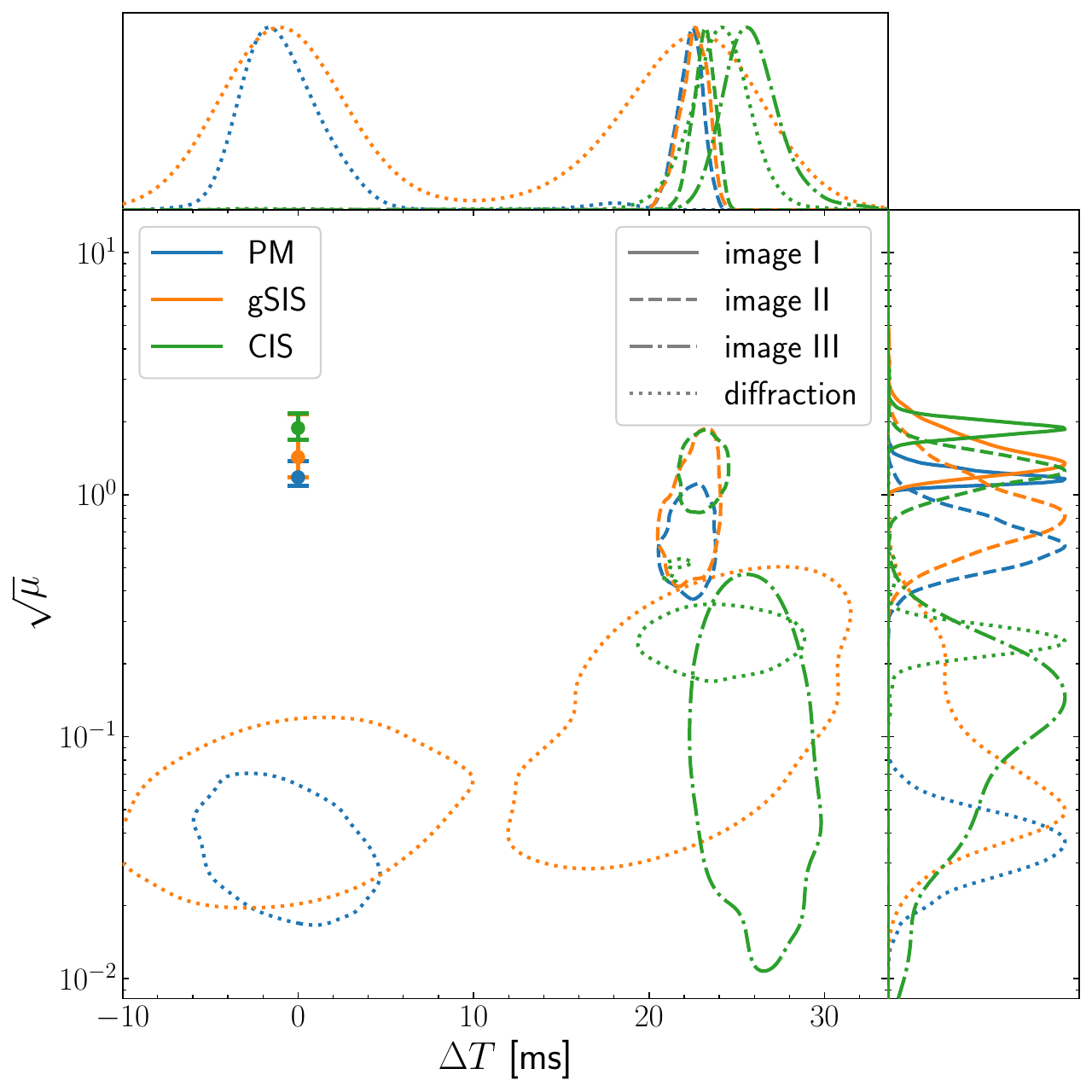}
    \caption{The posterior $90\%$ credible region of the magnification $\mu$ and the time delay $\Delta T$ relative to image I of each component in the waveform for each lens model.
    We plot $\sqrt{\mu}$ because the strain amplitude scales with the square root of the magnification.
    Both quantities for all images are computed in the geometrical optics limit from the lens parameter posteriors.
    The diffraction component is defined to be the residual of the waveform after subtracting the geometrical optics images, and its $\sqrt{\mu}$ and $\Delta T$ are defined to be the amplitude and time delay at its time of maximum strain amplitude.
    The posteriors for image I are plotted as circles with error bars because $\Delta T \equiv 0$.
    }
    \label{fig:mag_delta_T}
\end{figure*}

When fitting the data with a diffraction-lensed model, the waveforms can be approximated by a linear combination of geometrical optics images.
In Fig.~\ref{fig:images}, we show each image individually in the time domain.
We also show the diffraction component, which is just the residual after subtracting geometrical optics images from the full waveform.
In Fig.~\ref{fig:mag_delta_T} we show the absolute magnification $\mu$ and relative time delay $\Delta T$ (measured from the arrival time of the first image) of each image.
We also show the effective $\mu$ and $\Delta T$ for the diffraction component, which is just the time delay and amplitude at the time it attains maximum amplitude.
The waveform is dominated by the first (type I) and the second (type II) images, and the diffraction contribution is small.
The two dominant images lie $\sim 22 \rm \, ms$ apart from each other, and the other components lie close to either image.
This agrees with other studies on the same event (e.g. Refs.~\cite{Goyal:2025eqo, Hu:2025lhv, Chan:2025kyu}).
Note that the time delay coincides with the instantaneous period of the unlensed waveform at the peak of the merger~\cite{Chan:2025kyu}.
Heuristically, this could suggest that the feature being fitted away by the lensed model could actually be a feature of the intrinsic waveform itself, instead of a fine tuned lens set up that gives a time delay that coincides with the waveform period.
This is not an argument we pursue here but could be explored in future work.

\begin{figure*}
    \centering
    \includegraphics[width=0.48\linewidth]{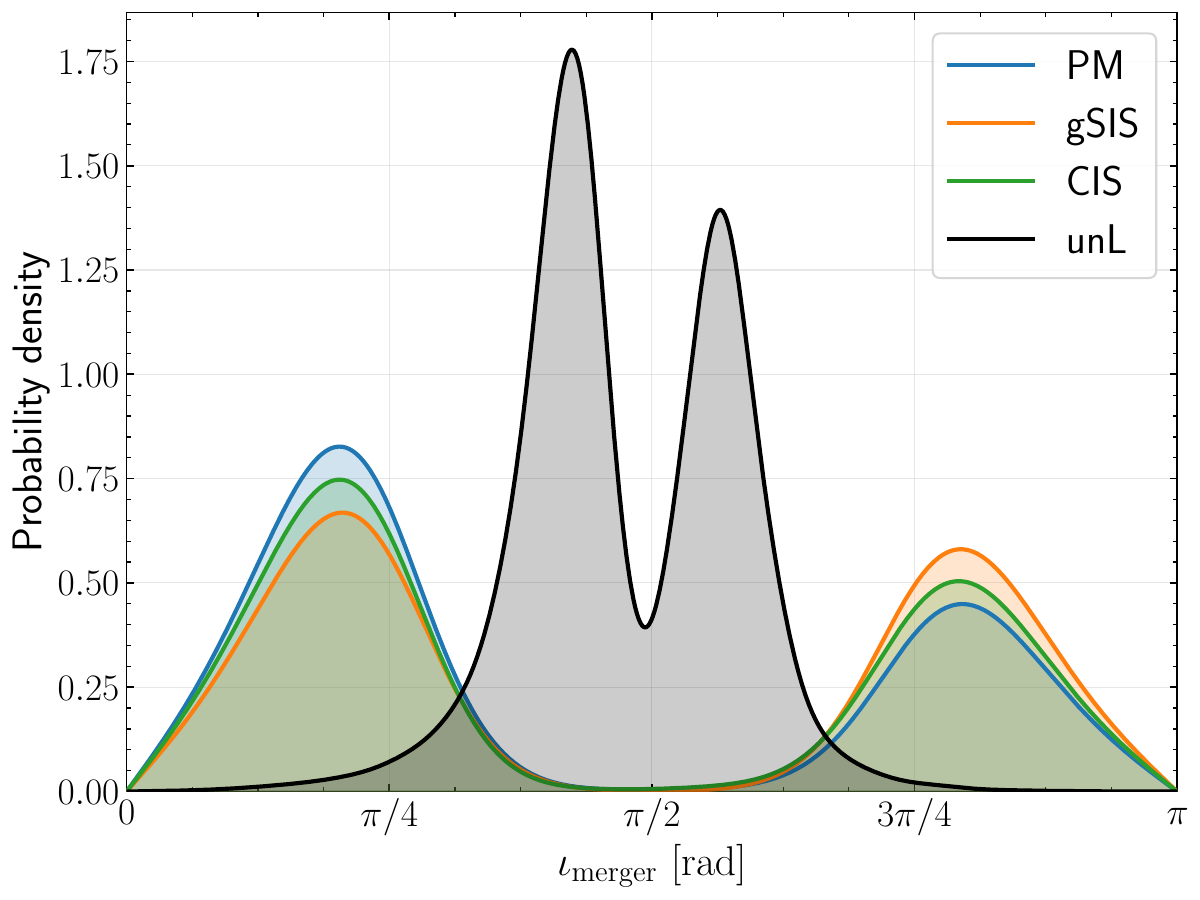}
    \includegraphics[width=0.96\linewidth]{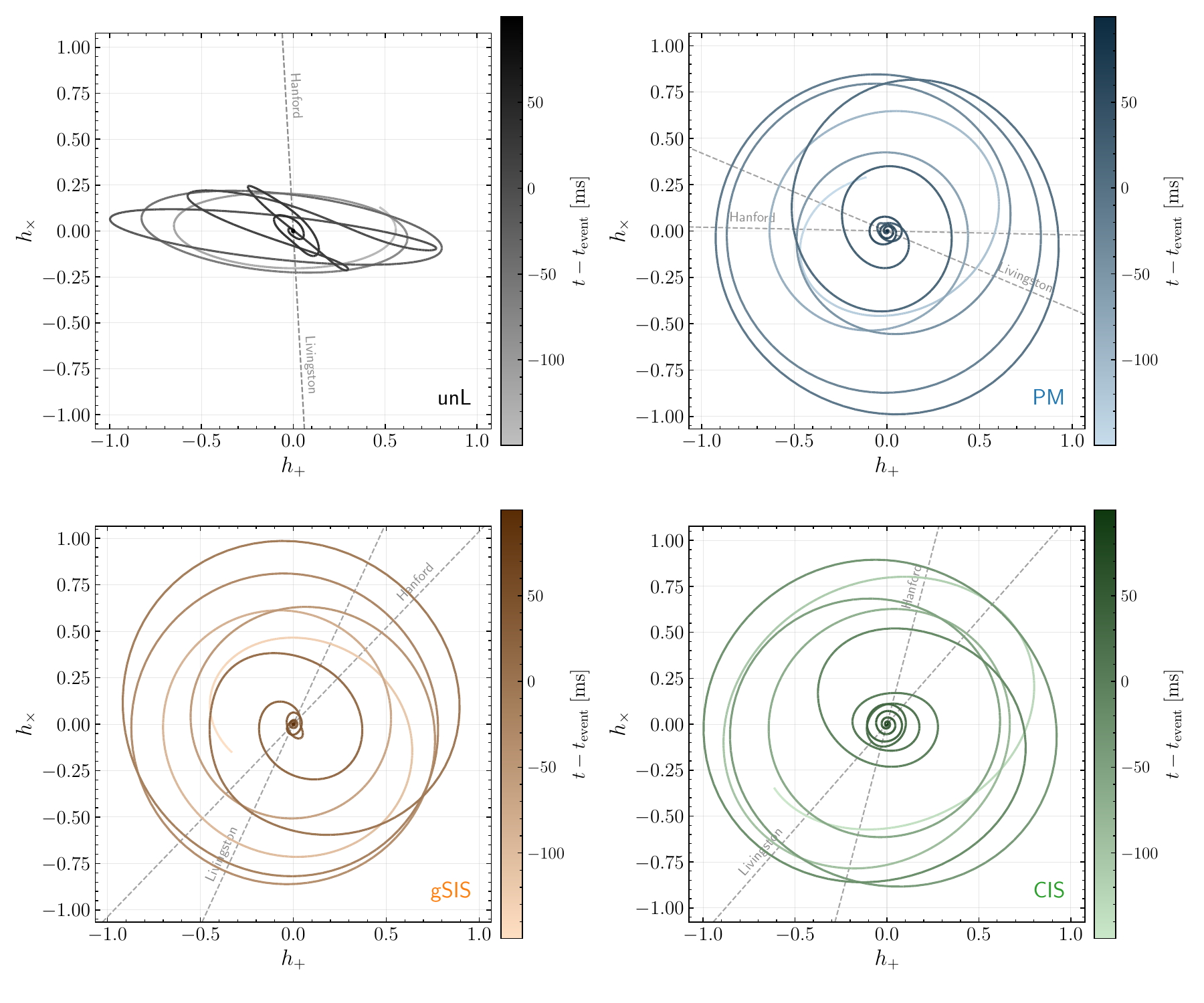}
    \caption{
    \textit{Top panel:} Posterior of the inclination $\iota_{\rm merger}$ of the orbital plane defined at the time of merger.
    The unlensed model gives an edge-on system while the lensed systems are more face-on.
    \textit{Bottom panels:} The time domain evolution of the maximum likelihood waveforms in the $h_+$ vs $h_\times$ plane.
    Because the unlensed waveform is edge-on, the phasor traces an ellipse with time varying size in the plane, and the ellipse rotates because the system is highly precessing.
    The same effect is not seen for the lensed models because the systems are more face-on, making the polarization ellipse close to circular.
    To obtain the detector strain, the waveforms are projected onto the gray dashed lines and scaled according to the orientation of each detector.
    The orientation of the lines is determined by the sky location and polarization angle of the maximum likelihood waveform plotted.}
    \label{fig:h+vshx}
\end{figure*}

A less appreciated fact about GW231123 is that when fit by an unlensed model, its orientation is close to edge-on at the time of merger.
Usually, the orientation of the binary is defined at a reference frequency $f_{\rm ref}$. 
To obtain the orientation at the merger, i.e.\ time of maximum strain amplitude $|h|$, we use the quaternion time series included in the \texttt{NRSur7dq4} model that encodes the rotation from the coprecessing frame to the inertial frame.
We can then define the inclination at merger $\iota_{\rm merger}$ as the angle between the orbital angular momentum $\vec{L}$ and the line of sight at the time of merger.
As shown in the top panel of Fig.~\ref{fig:h+vshx}, for the unlensed model, the posterior for $\iota_{\rm merger}$ is $\sim \pi/2$, which is consistent with an edge-on merger.
For the lensed models, the mergers are more face-on.

For face-on nonprecessing systems, in the $h_+$ vs $h_\times$ plane, the gravitational waveform phasor would trace a circle with time varying radius.
For edge-on ones, the phasor would trace an ellipse with time varying size, with the major axis aligned with the $h_\times = 0$ line due to the usual definition of $h_+$ and $h_\times$.
Therefore, at the same luminosity distance, edge-on mergers generally are not as loud as face-on ones because of a suppressed $h_\times$.
For highly precessing systems, the orbital plane precesses significantly, and for edge-on systems this shows up as a rotation in time of the polarization ellipse in the $h_+$ vs $h_\times$ plane~\cite{OShaughnessy:2012iol}.
Such an effect is apparent in the unlensed model panel in Fig.~\ref{fig:h+vshx}, and it does not show up significantly for the lensed models, because those systems are closer to face-on and the polarization is closer to circular.
This is closely related to the fact that precession is easier measured for edge-on systems~\cite{OShaughnessy:2012iol,Miller:2025eak}.

GW231123 was studied due to the extraordinary nature of its intrinsic parameters.
Heuristically, it seems unlikely that an event isolated for this reason would also be close to edge-on, when face-on mergers would have a higher detection volume.
This is suggestive of potential pathologies with fitting the GW231123 data with a pure unlensed waveform.
Due to the rotating polarization ellipse, the structure of edge-on highly precessing waveforms is richer than face-on ones, so the waveform seems to have the flexibility to fit away components in the data, e.g. glitches, lensing effects, or beyond-GR effects.
In this work, we have explored the lensing case.

The above observations are a part of an ongoing work by the authors, and are presented here because they seem to be relevant to this work. 
While suggestive, we make no claims that the arguments made here are precise or conclusive.

\section{NFW fit functions}
\label{app:nfw_fits}

\begin{figure*}
    \centering
    \includegraphics[width=0.99\linewidth]{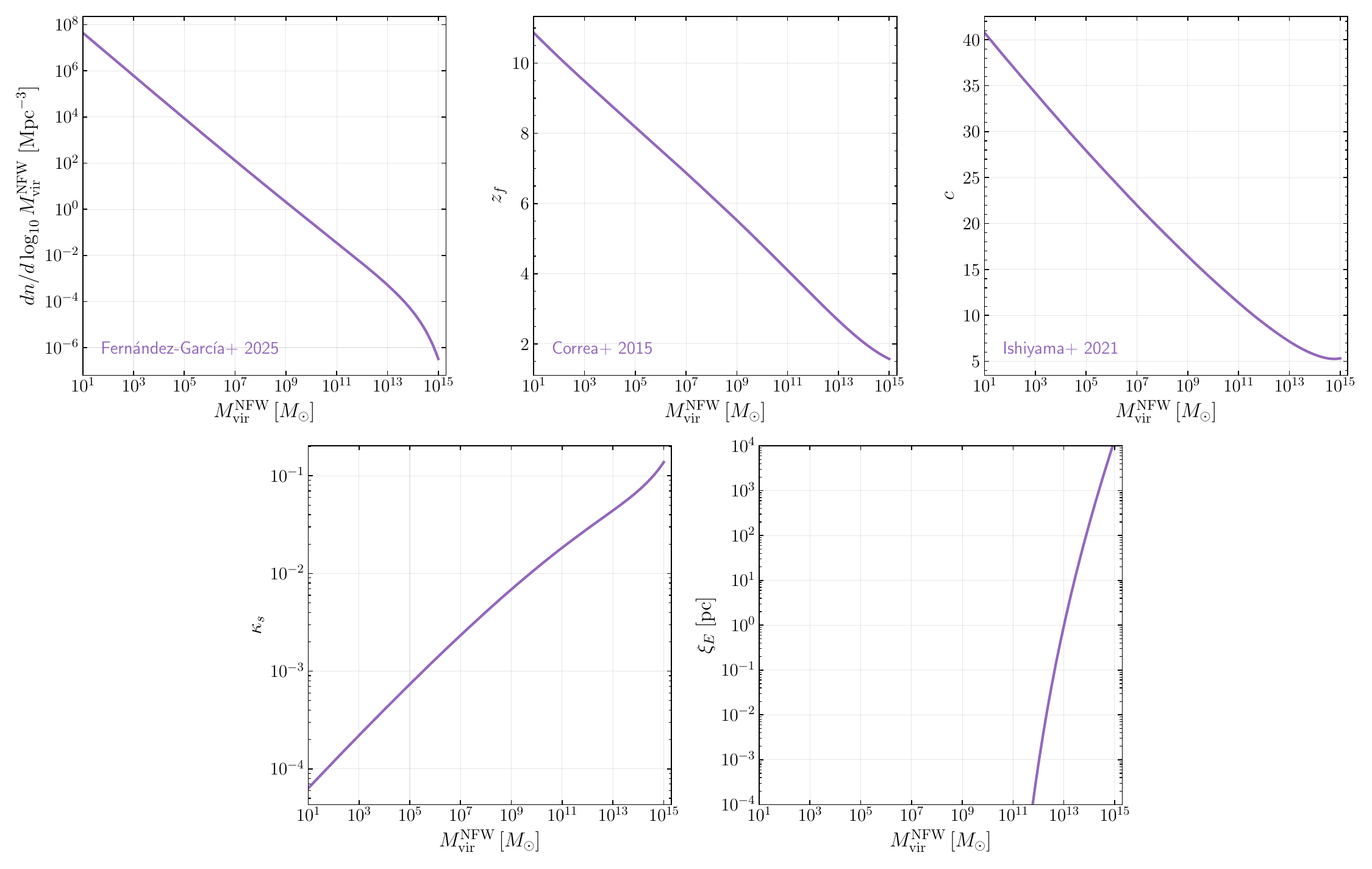}
    \caption{
    Quantities used in the main text that depend on the NFW halo virial mass $M^{\rm NFW}_{\rm vir}$, all evaluated at the lens redshift $z_L = 0.4$ (and, for $\kappa_s$ and $\xi_E$, source redshift $z_S = 0.8$): the halo mass function (top left, Ref.~\cite{Fernandez-Garcia:2025vzt}), the redshift of formation $z_f$ of the halos (top center, Ref.~\cite{Correa:2015kia}), the concentration parameter $c$ (top right, Ref.~\cite{Ishiyama:2020vao}), the characteristic convergence $\kappa_s$ of the halo (bottom left) and the Einstein radius $\xi_E$ (bottom right).
    }
    \label{fig:nfw_fits}
\end{figure*}

Several quantities used in the main text depend on the properties of the underlying NFW halo through fitting functions calibrated to cosmological $N$-body simulations, which we summarize here and show in Fig.~\ref{fig:nfw_fits}.
For the halo mass function $dn/dM^{\rm NFW}_{\rm vir}$, we use the mass function of Ref.~\cite{Fernandez-Garcia:2025vzt} as implemented in the \texttt{Colossus} package~\cite{Diemer:2017bwl}, adopting $M_{\rm vir} = M_{200m}$.
The concentration $c(M^{\rm NFW}_{\rm vir}, z)$ is taken from the fits of Ref.~\cite{Ishiyama:2020vao}, and the formation redshift $z_f$ from the analytic mass-accretion-history model of Ref.~\cite{Correa:2015kia}.
From $c$ and $M^{\rm NFW}_{\rm vir}$, we obtain $\rho_s$ and $r_s$ through Eqs.~\eqref{eq:NFW_c_rs} to~\eqref{eq:NFW_delta_c}, and hence the characteristic convergence $\kappa_s$ (Eq.~\eqref{eq:kappa_s}) and the true Einstein radius $\xi_E$, the latter obtained by solving $\bar{\kappa}(x_E) = 1$ with the NFW mean convergence of Ref.~\cite{Wright:1999jc}.
These are the ingredients used to convert between $M^{\rm NFW}_{\rm vir}$ and $M_{Lz}$ and to compute the optical depths in Sec.~\ref{sec:astro_prior}.

\bibliography{GW231123_lensing}

\end{document}